\documentclass[10pt,journal,compsoc]{IEEEtran}
\usepackage{amssymb}
\usepackage{amsbsy}
\usepackage{footnote}
\usepackage{amsmath}
\usepackage{array}
\usepackage{bbding}
\usepackage{amsfonts}
\makeatletter
\let\NAT@parse\undefined
\makeatother
\usepackage[colorlinks,linkcolor=blue,citecolor=blue]{hyperref}
\usepackage[usenames,dvipsnames]{color}
\usepackage{colortbl}
\usepackage{multirow}
\usepackage{booktabs}
\usepackage{makecell}

\definecolor{gray1}{gray}{.3}
\definecolor{gray2}{gray}{.5}
\definecolor{gray3}{gray}{.7}

\ifCLASSOPTIONcompsoc
  \usepackage[nocompress]{cite}
\else
  \usepackage{cite}
\fi
\ifCLASSINFOpdf
   \usepackage[pdftex]{graphicx}
   \usepackage{subfigure}
\else
\fi
\begin{document}
%
\title{A Survey on Role-Oriented Network Embedding}
%
%
%
%

\author{
       Pengfei~Jiao,
       Xuan~Guo,
       Ting~Pan,
        Wang~Zhang,
        and 
Yulong Pei

\thanks{Pengfei Jiao is with the College of Intelligence and Computing, Center of Biosafety Research and Strategy, Tianjin University, Tianjin, 300350, China (email: pjiao@tju.edu.cn).}
\thanks{Xuan Guo, Ting Pan and Wang Zhang are with the College of Intelligence and Computing, Tianjin University, Tianjin, 300350, China (email: guoxuan@tju.edu.cn; tingpan@tju.edu.cn; wangzhang@tju.edu.cn).}
\thanks{Yulong Pei is with the the Department of Mathematics and Computer Science, Eindhoven University of Technology, Eindhoven, 5600MB, the Netherlands (email: y.pei.1@tue.nl).}
 
}
%
%


\markboth{Journal of \LaTeX\ Class Files,~Vol.~14, No.~8, August~2015}%
{Shell \MakeLowercase{\textit{et al.}}: Bare Demo of IEEEtran.cls for Computer Society Journals}

%



\IEEEtitleabstractindextext{%
\begin{abstract}
Recently, Network Embedding (NE) has become one of the most attractive research topics in machine learning and data mining. NE approaches have achieved promising performance in various of graph mining tasks including link prediction and node clustering and classification.
A wide variety of NE methods focus on the proximity of networks. They learn community-oriented embedding for each node, where the corresponding representations are similar if two nodes are closer to each other in the network. 
Meanwhile, there is another type of structural similarity, i.e., role-based similarity, which is usually complementary and completely different from the proximity. In order to preserve the role-based structural similarity, the problem of role-oriented NE is raised.
However, compared to community-oriented NE problem, there are only a few role-oriented embedding approaches proposed recently.
Although less explored, considering the importance of roles in analyzing networks and many applications that role-oriented NE can shed light on, it is necessary and timely to provide a comprehensive overview of existing role-oriented NE methods.
In this review, we first clarify the differences between community-oriented and role-oriented network embedding. Afterwards, we propose a general framework for understanding role-oriented NE and a two-level categorization to better classify existing methods.
Then, we select some representative methods according to the proposed categorization and briefly introduce them by discussing their motivation, development and differences.
Moreover, we conduct comprehensive experiments to empirically evaluate these methods on a variety of role-related tasks including node classification and clustering (role discovery), top-k similarity search and visualization using some widely used synthetic and real-world datasets.
Finally, we further discuss the research trend of role-oriented NE from the perspective of applications and point out some potential future directions. The source code and datasets used in the experiments are available at \href{https://github.com/cspjiao/RONE}{Github}.
\end{abstract}

\begin{IEEEkeywords}
Network Embedding, Role Discovery, Structural Similarity, Unsupervised Learning, Experimental Analysis.
\end{IEEEkeywords}}

\maketitle

\IEEEdisplaynontitleabstractindextext

%
\IEEEpeerreviewmaketitle
\section{Introduction}\label{sec:Introduction}
\IEEEPARstart{N}{etwork} or graph is usually used to model the complex interaction relations in real-world data and systems~\cite{strogatz2001exploring,gao2016universal}, e.g., transportation, social pattern, cooperative behavior and metabolic phenomenon. By convention, the network (or graph) is usually abstracted as some nodes and their complicated and elusive links. To understand such data, network analysis can help to explore the organization, analyze the structure, predict the missing links and control the dynamics in complex systems. For a long time, researchers have proposed specially designed methods and models for different graph mining tasks, such as preference mechanism, hierarchical structure and latent space model for link predication~\cite{martinez2016survey}; grouping or aggregation, bit compression and influence based for network summarization~\cite{liu2018graph}; generalized threshold model, independent cascade model and linear Influence Model for information diffusion~\cite{zhang2016dynamics}. Among the core issues and applications of network analysis and graph mining, clustering~\cite{serrano2006clustering}, dividing the nodes into distinct or overlapping groups, has attracted attracted the most interest from different domains including machine learning and complex networks.

The field of network clustering has two main branches: \textbf{community detection}~\cite{girvan2002community,rosvall2007information,8531771} and \textbf{role discovery}~\cite{rossi2014role}. Community detection, the currently dominant clustering branch, is devoted to find common groups in which nodes interact more intensively than outside~\cite{fortunato2010community}. However, role discovery, which has a long research history in sociology~\cite{merton1968social} but had been inconspicuous in network science, groups the nodes based on the similarity of their structural patterns\cite{liu2012social}, such as the bridge or hub nodes~\cite{gilpin2013guided}. In general, nodes in the same community are likely to be connected to each other, while nodes in the same role may be unconnected and often far away form each other. Since their rules on dividing nodes are fundamentally different, the two branches are usually considered as orthogonal problems. A variety of algorithms and models are proposed for both of the two branches. For community detection, the modularity optimization~\cite{chen2014community,li2013multicomm}, statistical model~\cite{stanley2016clustering,7745890}, non-negative matrix factorization~\cite{yang2014unified,pei2015nonnegative,ma2017evolutionary,ma2018community} and deep learning methods~\cite{tu2018unified,liu2020deep} are developed and show crucial influence for other tasks and applications, such as recommendation\cite{feng2015personalized,zheng2019personalized} and identifying criminal gangs~\cite{lu2014algorithms}. Some surveys on community detection can be seen in~\cite{fortunato2010community,fortunato2016community,jin2021survey}. For role discovery, traditional methods are usually graph based and related to some equivalence, such as the structural~\cite{lorrain1971structural}, regular~\cite{white1983graph} and stochastic equivalence~\cite{holland1981exponential,nowicki2001estimation}. Blockmodels\cite{faust1992blockmodels} and mixed-membership stochastic blockmodels~\cite{airoldi2008mixed} are the important and influential methods are based on the graph. Besides, there are also some combinatorial or heuristic methods \cite{arockiasamy2016combinatorial} for this problem.

 \begin{figure*}
    \centering
    \subfigure[Truth label]{
    \includegraphics[width=.3\textwidth]{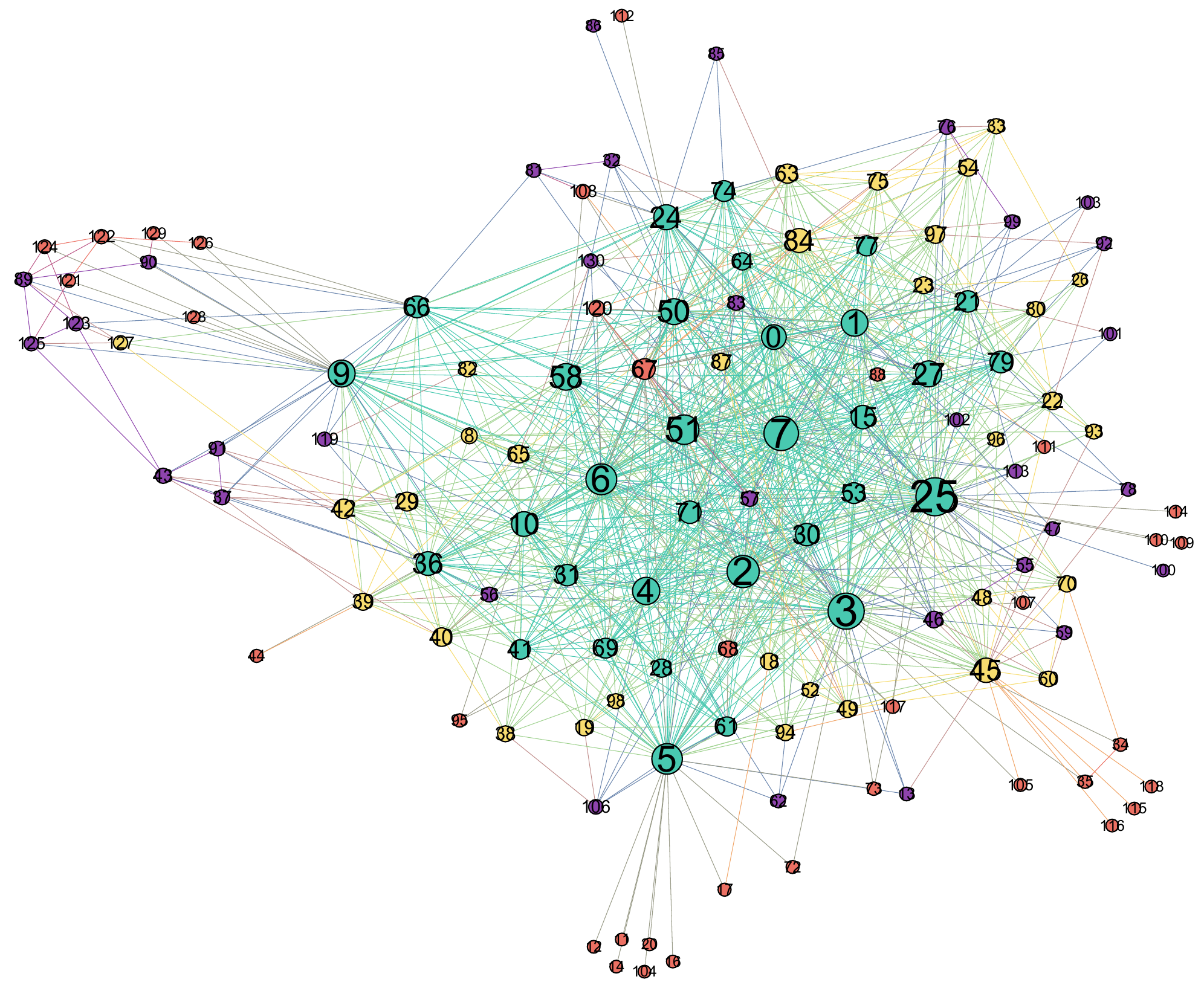}
    }
    \subfigure[Community detection]{
    \includegraphics[width=.3\textwidth]{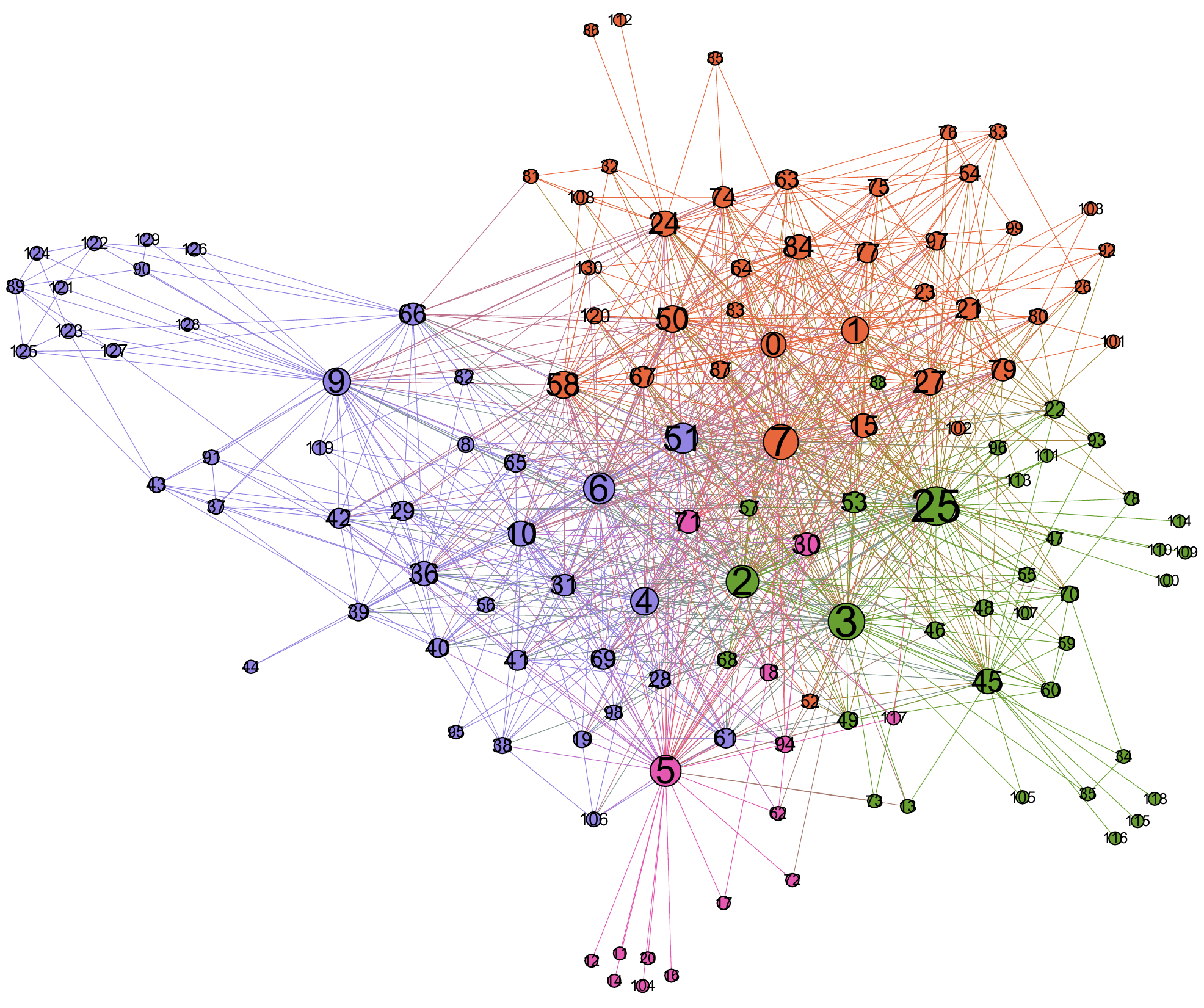}
    }
    \subfigure[Role discovery]{
    \includegraphics[width=.3\textwidth]{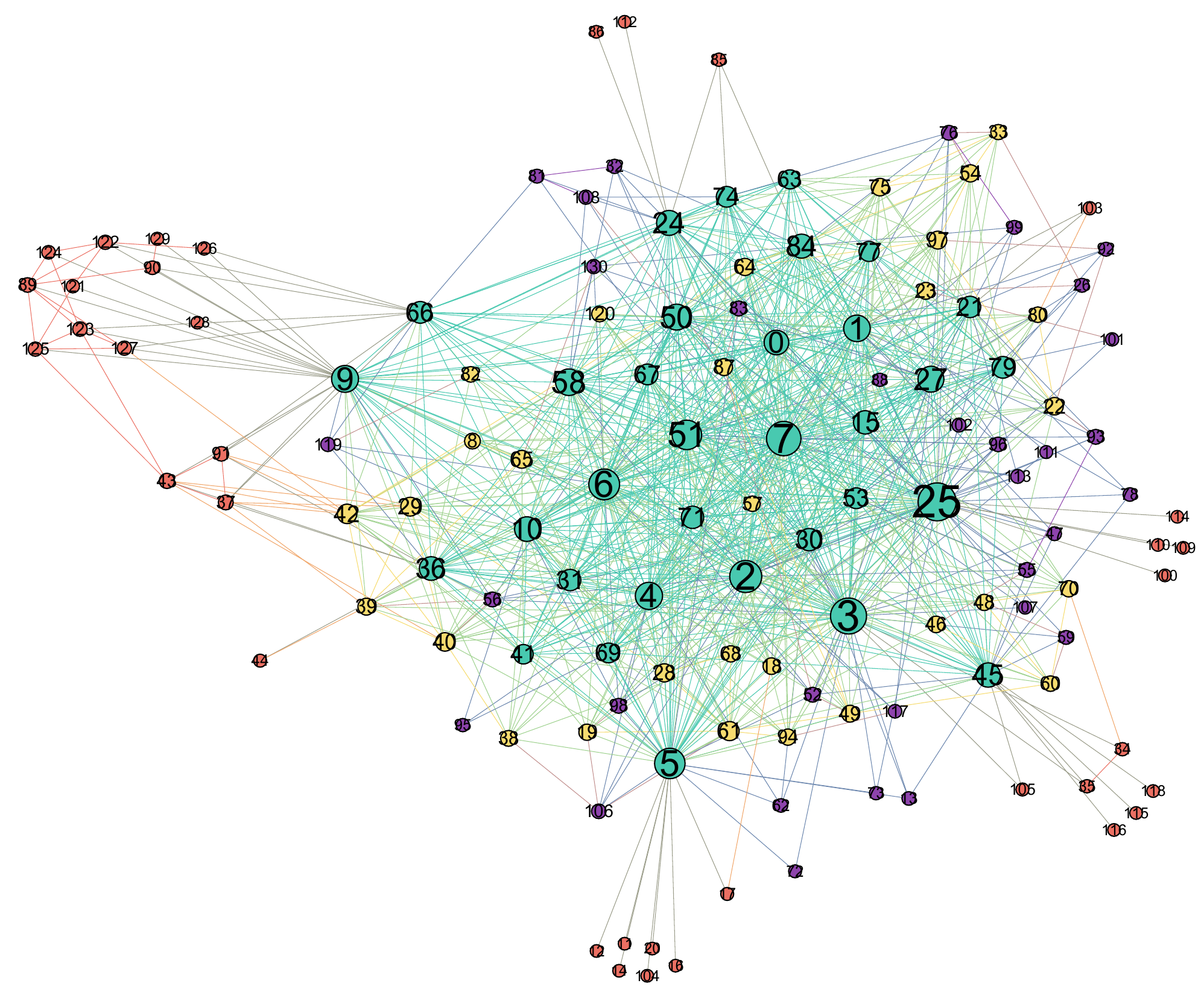}
    }
    \caption{Air Brazil network for understanding the community detection and role discovery in network clustering. Nodes with same clustering label have same color. (a) Brazil network with ground-truth clustering label; (b) Community detection with Louvain~\cite{blondel2008fast}; (c) Role discovery with RolX~\cite{henderson2012rolx}.}
    \label{fig:Brazil-sample}
\end{figure*}

Here we take the Brazilian air-traffic network as an example. As shown in Fig.~\ref{fig:Brazil-sample} (a), the nodes and edges denote the airports and their direct flights, respectively. The clustering labels are marked based on the activity of nodes~\cite{ribeiro2017struc2vec}. The size and color of the circle represent the degree and label of the node, respectively. It can be observed that, the nodes with the same label (color), i.e., they are structurally similar, are usually not connected.
To better illustrate these two clustering tasks, we choose two typical methods, Louvain~\cite{blondel2008fast} and RolX~\cite{henderson2012rolx}, which are specially designed for community detection and role discovery, respectively. The clustering results are presented in Fig.~\ref{fig:Brazil-sample} (b) and (c). Community detection divides this network into some tightly connected groups, which means the airports with more flights among them belong to the same community. However, the clustering result detected by the role discovery, usually related to the flow and scale of the airport, is closer to the truth. 


In recent years, network embedding (NE) has become has become the focus of studying graph structure and been demonstrated to achieve promising performance in many downstream tasks, e.g., node classification and link prediction. The motivation of NE is to transform the network data into independently distributed representations in a latent space and these representations are capable of preserving the topological structure and properties of the original network. On the whole, current methods for NE can be categorized into two types: shallow and deep learning (here we focus on unsupervised NE approaches without explicit mentioning).  The former includes the matrix factorization and random walk based methods. The goal of matrix factorization methods is to learn node embedding via the low-rank approximation and approaching the adjacency matrix or higher-order similarity of the network, such as the singular value decomposition, non-negative matrix factorization and NetSMF~\cite{qiu2019netsmf}. With the different random walk strategies, a series of methods have been proposed to optimize the co-occurrence probability of nodes and learn effective embeddings. The later is mainly rooted in autoencoder and graph convolutional networks. These methods generally consist of the encoder, similarity function and decoder. There are also some attributes, characteristics and constraints that can be combined to enhance the embedding, and VGAE, GAT and GraphSAGE are some representatives of such methods. There are also some other types of embedding approaches, e.g., the latent feature model, but discussion on general NE methods is out of our scope. We refer the interested readers to some recent survey papers on NE~\cite{cui2018survey,zhang2018network,zhang2020deep,wu2020comprehensive,goyal2018graph,cai2018comprehensive,hamilton2017representation,chen2018tutorial}.

However, most of these methods, whether or not for network clustering, are designed for modeling the \textbf{proximity}, i.e. the embedding vectors are community oriented. They fail to capture the structural similarity, or the role information~\cite{rossi2020proximity}, Therefore, it raises several inherent challenges in the research of role-oriented network embedding. Firstly, the most and important is that two nodes with structural similarity have nothing to do with their distance, which makes it difficult to define the loss function effectively. Secondly, strict role definitions, such as some definitions based on equivalence, are difficult to be implemented in real-world networks especially large-scale networks. Thirdly, the distribution of nodes with same role in the network is very complex and interaction patterns between different roles are unknown. 

In essence, there are still some scattered methods being proposed one after another recently years. These methods uses various embedding mechanisms. Struc2vec~\cite{ribeiro2017struc2vec} leverages random walks on graphs in which edges are weighted based on structural distances. DRNE develops a deep learning framework with layer-normalized~\cite{tu2018deep} LSTM model to learn regular equivalence. REACT~\cite{pei2019joint}, generating embeddings via matrix factorization, focus on capturing both community and role properties. Though the number of diverse role-oriented embedding methods is gradually increasing, there is still a lack of systematic understanding of role-oriented network embedding. Besides, we also lack a taxonomy for deep thinking of this problem. Meanwhile, there is short of performance and efficiency comparison of currently methods.  All these limit the development and applications of role embedding.

So in this survey, we systematically analyze role-oriented network embedding and the analysis can help to understand the internal mechanism of currently methods. First, we propose a two-level categorization scheme for existing methods which is based on embedding mechanism of currently methods and models. Further more, we evaluate selected embedding methods from the perspectives of both efficiency and effectiveness on different tasks related to role discovery. In specific, we conduct comprehensive experiments on some representative methods on running time (efficiency), node classification and clustering (role discovery), top-k similarity search and visualization with widely used benchmark networks. Last, we summarize the applications, challenges and future directions of role-oriented network embedding.

Some surveys on network embedding, community detection, role discovery, and deep learning on graph have been conducted. Our survey has several essential differences compared to these works.~\cite{fortunato2010community,fortunato2016community,jin2021survey} mainly study the problem of community detection with different focuses from the perspective of network analysis and machine learning. \cite{rossi2014role} is a seminal work in reviewing the development and methodology of role discovery. However, this survey is relatively outdated where more advanced methods, e.g., deep learning based methods, have not been discussed. Besides, these surveys focus on the methods specific for community or role task, while our work studies roles with a focus on network embedding approaches which can preserve the role information. The surveys~\cite{hamilton2017representation,cui2018survey,zhang2018network,chen2018tutorial} are influential works on network embedding from different principles. However, they all focus on community-oriented methodology. Similarly, some graph embedding\footnote{We do not distinguish the difference between network embedding and graph embedding.} reviews~\cite{goyal2018graph,cai2018comprehensive}, however, except for some technologies, have nothing to do with the role-oriented embedding. Meanwhile, surveys such as~\cite{wu2020comprehensive} and~\cite{zhang2020deep} introduce the effective deep learning framework and methods on graph or networks. They focus on general problems of how to use machine learning on networks and are less relevant to our problem. One relevant work is~\cite{rossi2020proximity}, it clarifies the difference between the community-oriented and role-oriented network embedding for the first time, and proposes the normal mechanisms which can help to understand if a method is designed for community or role. However, it does not systematically discuss the series of role-oriented NE methods: some advanced methods have been ignored, some introduced works are not used for role discovery or role related tasks. Moreover, it does not evaluate methods empirically by analyzing the relevant data, tasks and performance. Another recent work~\cite{jin2021towards}, which introduces some structural node embedding methods and evaluate them empirically, is the most relevant to our work. In analysis, they mainly focus on analyzing the relationships between NE methods and equivalence. In evaluation, they evaluate the discovered roles on direct tasks such as role classification and clustering. In contrast, we concentrate on analyzing advantages and disadvantages of different role-oriented approaches using a new two-level categorization from the analysis perspective. We conduct more comprehensive experiments to evaluate different methods w.r.t. both efficiency and effectiveness in role discovery and downstream tasks including running time, classification, clustering, visualization, and top-k similarity search.

To sum up, our survey has several contributions as follows.
\begin{itemize}
    \item We first show the summary of role-oriented network embedding and discuss the relationship and differences of it and community oriented.
    \item We propose a two levels categorization schema of currently role-oriented embedding methods and briefly describe their formalization, mechanism, task, connection and difference.
    \item We provide full experiments of popular methods of each type on different role-oriented tasks and detailed comparison on effectiveness and efficiency.
    \item We share all the open-source code and widely used network datasets on Github \url{} and point out the development and questions of role-oriented network embedding.
\end{itemize}

\section{Notations and Framework}\label{sec:definition}

In this section, we give formal definitions of basic graph concepts and role-oriented network embeddings. In Table~\ref{tab.notation}, we summarize the main notations used throughout this paper. Then, we propose a unified framework for understanding the process of role oriented network embedding.

\newtheorem{definition}{Definition}
\begin{definition}[Network]
A network is denoted as $\mathcal{G}=(\mathcal{V},\mathcal{E})$, where $\mathcal{V} =\{v_1,...,v_n \}$ is the set of $n$ nodes and $\mathcal{E} \subseteq \mathcal{V} \times \mathcal{V}$ is the set of edges. An edge $e_{ij} = (v_i, v_j) \in \mathcal{E}$ denotes the link between node $v_i$ and $v_j$.
\end{definition}

 In usual, a network $\mathcal{G}$ is represented by an weight matrix $\mathbf{A} \in \mathbb{R}^{n \times n}$. If $e_{ij} \in \mathcal{E}$, $\mathbf{A}_{ij} > 0$ ($\mathbf{A}_{ij} = 1$ for an unweighted network and $\mathbf{A}_{ij} = \mathbf{A}_{ji}$ for an undirected network), otherwise $\mathbf{A}_{ij} = 0$. Some networks may have an attribute matrix $\mathbf{X} \in \mathbb{R}^{n \times x}$ whose $i$th row represents attributes of $v_i$.
 For an undirected network, denote the degree of node $v_i$ as $d_i = \sum_{j}\mathbf{A}_{ij}
 $, and we have the degree matrix $\mathbf{D}=\mathrm{Diag}(d_1,...,d_n)$. $\mathbf{L}=\mathbf{D}-\mathbf{A}$ is called the Laplacian matrix, it can be decomposed as $\mathbf{L}=\mathbf{U} \Lambda \mathbf{U}^\top$ where $\Lambda = \mathrm{Diag}(\lambda_1,...,\lambda_n)$ is the matrix of eigenvalues satisfying $\lambda_1 \le \lambda_2 \le ... \le \lambda_n$. 
 
 Denote the $k$-hop ($k>0$) reachable neighbor set of node $v_i$ as $\mathcal{N}^k_
i$ ($k$ is omitted when $k=1$), where the shortest path between $v_i$ and each node $v_j \in \mathcal{N}^k_
i$ is less than or equal to $k$. For a directed network, use $d_i^+$, $d_i^-$, $\mathcal{N}^{k+}_i$ and $\mathcal{N}^{k-}_i$ to represent the out/in-degree and $k$-hop reachable out/in-neighborhood of $v_i$ respectively. Unless otherwise stated, a model is discussed on unweighted undirected networks without attributes in later part of this paper. 

\begin{table}[!t]
\renewcommand{\arraystretch}{1.7}
\caption{Main Notations.}

\label{tab.notation}
\centering
\begin{tabular}{c|c}
\hline
\bfseries Notation & \bfseries Definition\\
\hline
$\mathcal{G}=(\mathcal{V},\mathcal{E})$ & the network/graph with node set $\mathcal{V}$ and edge set $\mathcal{E}$\\
\hline
$\mathcal{N}^k_i$& the  set of $v_i$'s $k$-hop reachable neighbors\\
\hline
$\mathcal{G}^k_i$& the subgraph induced by $v_i$ and $\mathcal{N}^k_
i$\\
\hline
$d_i$& the degree of node $v_i$\\
\hline
$s_{ij}$& the shortest path between $v_i$ and $v_j$\\
\hline
$\mathbf{X}$& attribute matrix \\
\hline
$\mathbf{I}$& identity matrix\\
\hline
$\mathbf{H}$& embedding matrix\\
\hline
$\mathbf{F}_{\rm m}$& the feature matrix extracted by or in method $\rm m$\\
\hline
$\mathbf{S}_{\rm m}$& the similarity matrix obtained by or in method $\rm m$\\
\hline
$\circ$, $\langle \langle \cdot \rangle \rangle$& the concatenation operator\\
\hline
\end{tabular}
\\\footnotesize{*For conveniece, method notation $\rm m$ is omitted in some descriptions.}
\end{table}

\begin{figure*}
    \centering
    {\includegraphics[width=0.9\linewidth]{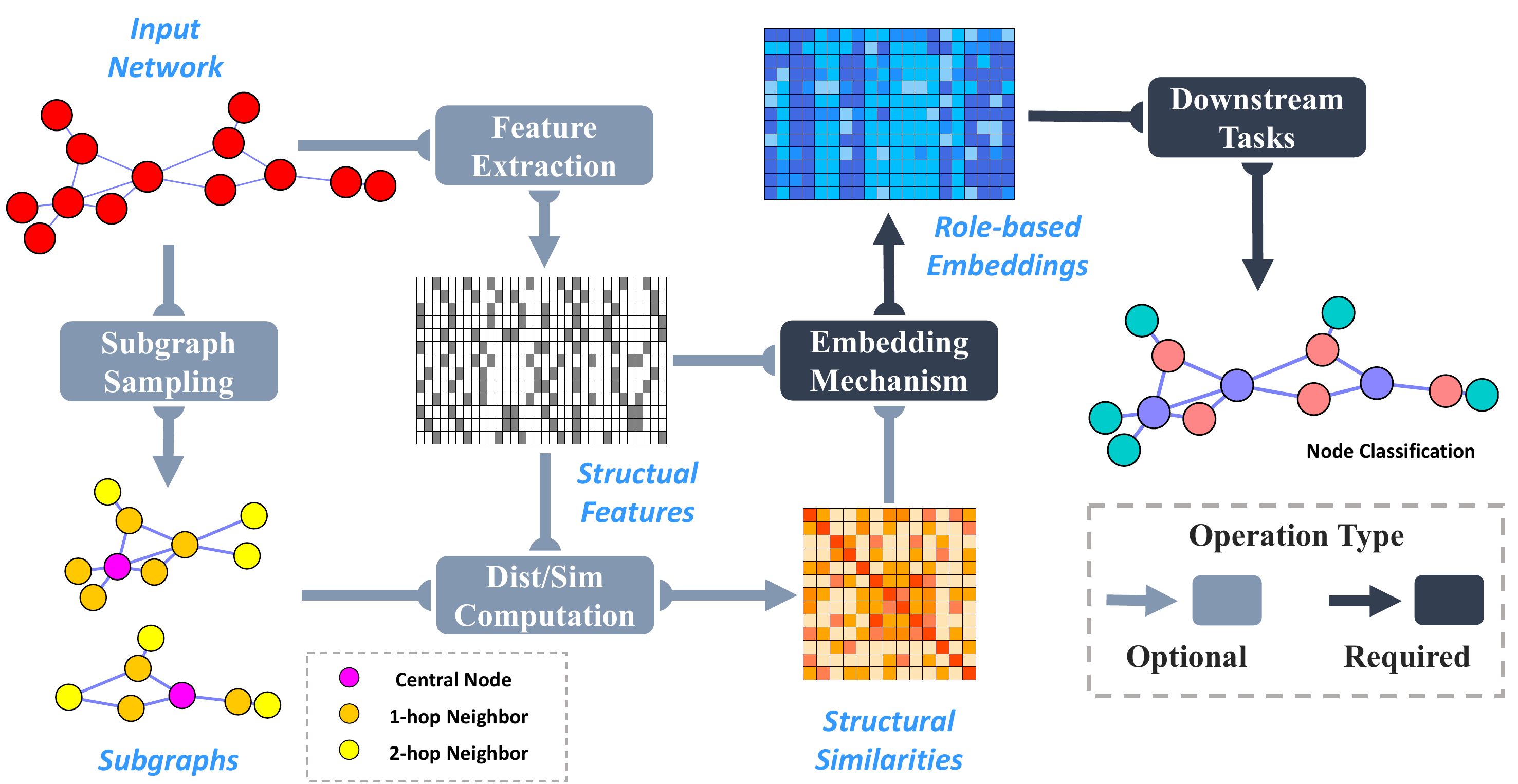}}
    \caption{The common framework of role-oriented network embedding methods includes two main step: structural property extraction and embedding. The former one can be accomplished via a variety of ways of which some are feature-based and some are similarity-based methods. On the extracted properties, role-oriented embedding methods then employ some specific embedding mechanisms to generate embeddings. Note that the discussed role-oriented methods are unsupervised. Thus, though the generated embeddings can be applied on some downstream tasks with ground truth, the whole process of embedding generation has no interaction with the target tasks.}
\label{fig:paradigm}
\end{figure*}

\begin{definition}[Motif/Graphlet]
A motif/graphlet $\mathcal{M} = (\mathcal{V}_{\mathcal{M}},\mathcal{E}_{\mathcal{M}})$ is a small connected subgraph representing particular patterns of edges on several nodes. The pattern can be repeated in or across networks, i.e., many subgraphs can be sampled from networks and isomorphic to it. Nodes automorphic to each other, i.e., having the same connectivity patterns, are in the same orbits. 
\end{definition}

\begin{figure}
    \centering
    {\includegraphics[width=0.8\columnwidth]{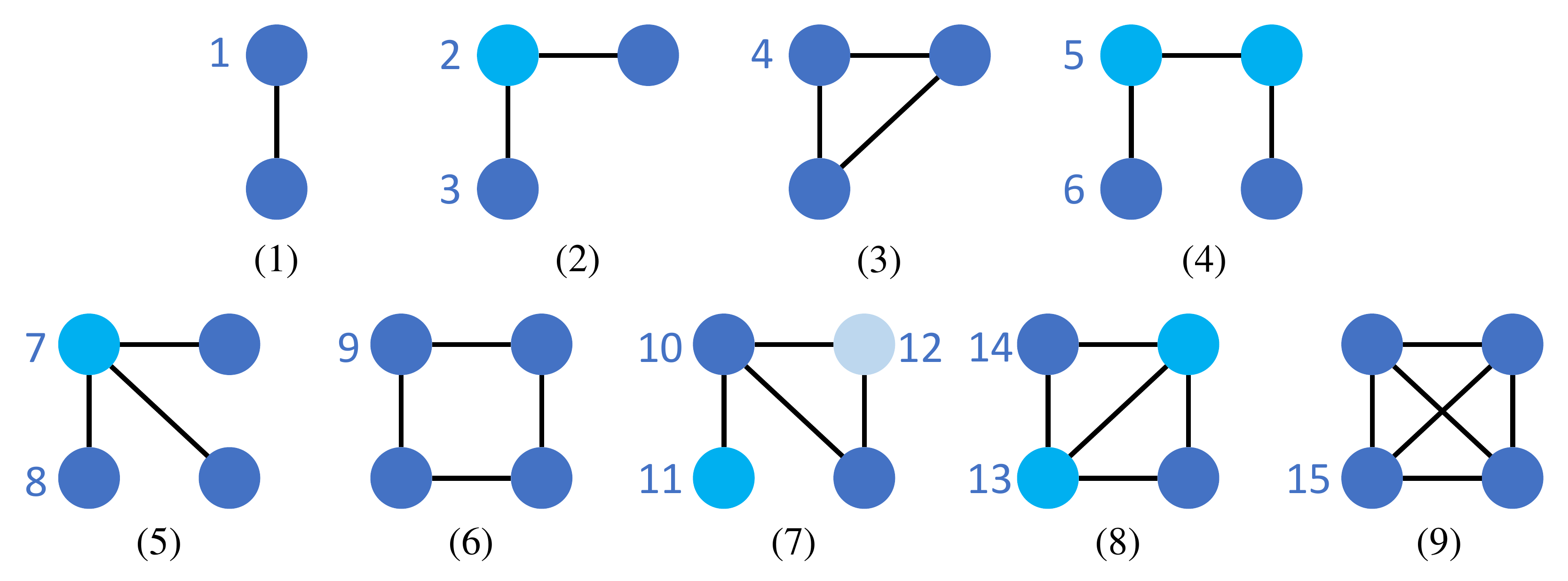}}
    \caption{Motifs and orbits (denoted by numbers) with size 2-4 nodes.}
\label{fig:motifs}
\end{figure}

For unweighted networks, there are 9 motifs and 15 orbits with size 2-4 nodes as shown in Fig.~\ref{fig:motifs}. Because of their ability to model the smallest but most fundamental connectivity patterns, motifs are wildly used for capturing structural similarities and discovering roles.

\begin{definition}[Network Clustering]
A clustering $\boldsymbol{\mathcal{R}}
=\{ \mathcal{R}_1,...,\mathcal{R}_k\}$ of network $\mathcal{G}=(\mathcal{V},\mathcal{E})$ is a group of node sets satisfying $\forall 1 \le i \le k: \mathcal{R}_i \neq \emptyset, \mathcal{R}_i \subset \mathcal{V}$ and $\bigcup_1^{k} \mathcal{R}_i = \mathcal{V}$. 
In this paper, we discuss about hard clustering, i.e., $\forall 1 \le i < j \le k: \mathcal{R}_i \cap \mathcal{R}_j = \emptyset$. If $\mathcal{R}_i \cup \mathcal{R}_j \neq \emptyset$, it is usually called overlapping or soft clustering.
\end{definition}

For the \textbf{community detection}, each set $\mathcal{R}_i$ is a tightly interconnected collection of nodes.
And for \textbf{role discovery}, it usually composed of unconnected nodes which have similar structural patterns or functions. So every network clustering algorithm is committed to achieve the clustering results under different goal constraints. However, there is no common understanding of role equivalence or similarity, which leads to multifarious definitions of equivalence and designs of similarity computation. For example, two nodes are automorphic equivalent~\cite{autoandsto-equi} as the subgraphs of their neighborhood are isomorphic, while regular equivalence~\cite{regular-equi} means that if two nodes have the same roles, there neighbors have the same roles. 

\begin{definition}[Network Embedding]
Network Embedding is a process to map nodes of network $\mathcal{G}=(\mathcal{V},\mathcal{E})$ to low-dimensional embeddings $\mathbf{H} \in \mathbb{R}^{n \times r}$  so that $r \ll n$. In general, for nodes $v_i$ and $v_j$, if they are similar in the network, their embedding vectors $\mathbf{H}_i$ and $\mathbf{H}_j$ will be close in the low dimensional space. 
\end{definition}

With the node embedding, we can take it for different network tasks. If we focus on the community detection or link predication, we want to discriminate by embeddings whether the nodes are connected or likely to be connected. However, for role discovery, the embeddings should reflect some structural patterns including local properties like subgraph isomorphism, global properties like regular equivalence and higher-order properties like motifs.


Based on the discussion and notations above, here we firstly propose a unified framework for understanding the role-oriented network embedding. To our knowledge, it can cover almost all the existing methods and models in a unified way. 
The framework is illustrated in Fig. \ref{fig:paradigm}. Structure of networks is discrete, but embeddings are usually designed to lie in continuous space. Thus, role-oriented embedding methods always take two steps to capture structrual properties and generate embeddings respectively to bridge the gulf between two spaces: 
\begin{itemize}
    \item \textbf{Structure Property Extraction}. The ways to extract structural information are diverse. Some methods such as RolX~\cite{henderson2012rolx} and DRNE~\cite{tu2018deep} leverages some primary structural features including node degree, triangle numbers. Part of these methods such as SPINE~\cite{ijcai2019-333} will continue to transform the features into distances or similarities. There are also some methods captureing similarity between node-centric subgraphs. For example, struc2vec~\cite{ribeiro2017struc2vec} compute structural distances between $k$-hop subgraphs based on degree sequences of the subgraphs. SEGK~\cite{nikolentzos2019learning} employs one graph isomorphism test skill called graph kernel on subgraphs. As the result of these hand-craft process, these structural properties are contained in interim features or pair-wise similarities.
    \item \textbf{Embedding}. The extracted properties then are mapped into embedding space via different mechanisms. In the process of embedding, the structural properties are used as inputs or training  guidance. For example, RolX~\cite{henderson2012rolx} and SEGK~\cite{nikolentzos2019learning} apply low-rank matrix factorization on feature matrix and similarity matrix respectively, as they implicitly or explicitly reflect whether nodes are structurally similar. Struc2vec\cite{ribeiro2017struc2vec,ijcai2019-333} leverages word embedding methods to the similarity-biased random walks. DRNE~\cite{tu2018deep} utilizes LSTM~\cite{hochreiter1997long} on degree-ordered node embedding sequences to capture regular equivalence with a degree-guided regularizer. 
\end{itemize}

As the embeddings capture crucial structural properties, they can be used on the downstream tasks such as role-based node classification and visualization. With this framework, we generalize the process of role-oriented embedding. However, as we can learn from Fig. \ref{fig:paradigm}, the core of designing role-oriented embedding methods is the way to extract structural properties. In contrast, the embedding mechanisms for mapping structural features/similarities into low-dimensional continuous vector space are much more regular. Thus, We introduce the popular methods in the next section from the perspective of embedding mechanisms.


\section{Algorithm Taxonomy}\label{sec:algorithm}

In this section, we introduce these approaches categorized according to their embedding mechanisms. In detail, we propose a two-level classification ontology for these popular methods. Similar to the taxonomy of community oriented network embedding, we divide these into three categories, low-rank matrix factorization, random walk based and deep learning methods from the first level. Further more, with there embedding mechanisms and constraint information, we give a more refined classification taxonomy as shown in TABLE~\ref{tab.methodlist}. At the same time, we also list the tasks which can be served by different methods. Next, we will introduce these methods in detail.

\newcommand{\tabincell}[2]{\begin{tabular}
{@{}#1@{}}#2\end{tabular}}
\begin{table*}[htbp]
\renewcommand{\arraystretch}{1.35}
\caption{A summary of role-oriented embedding methods. The abbreviations of tasks CLF, CLT, LP, ER, NA, SS and Vis denote node
classification/clustering, link prediction, entity resolution/Network alignment/Top-k similarity search and visualization, respectively. }
\label{tab.methodlist}
\centering
\begin{tabular}{|c|c|c|p{1.3cm}<{\centering}|p{1.3cm}<{\centering}|p{1.3cm}<{\centering}|p{1.3cm}<{\centering}|c|}
\hline
\multirow{2}{*}{\bfseries Method} & \multicolumn{2}{c|}{\multirow{2}{*}{\bfseries Embedding Mechanism}} & \multicolumn{4}{c|}{\bfseries Conducted Tasks} & \multirow{2}{*}{\bfseries Year}\\ \cline{4-7}
~ & \multicolumn{2}{c|}{~} & \bfseries Vis & \bfseries CLF/CLT & \bfseries ER/NA/SS & \bfseries LP & \\ \hline

RolX\cite{henderson2012rolx} & \multirow{10}{*}{\tabincell{c}{low-rank \\ matrix \\ factorization \\ (Sec.\ref{sec:lowrank})}}& \multirow{5}{*}{\tabincell{c}{on structural feature \\ matrix (Sec.\ref{sec:lowrank-feature}) }} & \textcolor{blue}{\CheckmarkBold} & \textcolor{blue}{\CheckmarkBold} & \textcolor{red}{\XSolidBold} & \textcolor{red}{\XSolidBold} & 2012\\ \cline{1-1} \cline{4-8}
GLRD\cite{gilpin2013guided} &  &  & \textcolor{red}{\XSolidBold} & \textcolor{red}{\XSolidBold} & \textcolor{blue}{\CheckmarkBold} & \textcolor{red}{\XSolidBold} & 2013\\ \cline{1-1} \cline{4-8}
RID$\large\boldsymbol{\varepsilon}$Rs\cite{gupte2017role} &  &  & \textcolor{blue}{\CheckmarkBold} & \textcolor{blue}{\CheckmarkBold} & \textcolor{blue}{\CheckmarkBold}  & \textcolor{red}{\XSolidBold} & 2017\\ \cline{1-1} 
\cline{4-8}
GraphWave\cite{donnat2018learning} &  &  & \textcolor{blue}{\CheckmarkBold} & \textcolor{blue}{\CheckmarkBold} & \textcolor{red}{\XSolidBold} & \textcolor{red}{\XSolidBold} & 2018\\ \cline{1-1} \cline{4-8}
HONE\cite{rossi2020structural} &  &  & \textcolor{blue}{\CheckmarkBold} & \textcolor{red}{\XSolidBold}  & \textcolor{blue}{\CheckmarkBold} & \textcolor{blue}{\CheckmarkBold} & 2020\\ \cline{1-1} \cline{4-8} \cline{3-3}
xNetMF\cite{heimann2018regal} &  & \multirow{5}{*}{\tabincell{c}{on structural similarity \\ matrix  (Sec.\ref{sec:lowrank-similarity}) }} & \textcolor{red}{\XSolidBold} & \textcolor{red}{\XSolidBold} & \textcolor{blue}{\CheckmarkBold} & \textcolor{red}{\XSolidBold} & 2018\\\cline{1-1} \cline{4-8}
EMBER\cite{jin2019smart} &  &  & \textcolor{red}{\XSolidBold} & \textcolor{blue}{\CheckmarkBold}  & \textcolor{blue}{\CheckmarkBold} & \textcolor{red}{\XSolidBold} & 2019\\ \cline{1-1} \cline{4-8}
SEGK\cite{nikolentzos2019learning} &  &  & \textcolor{blue}{\CheckmarkBold} & \textcolor{blue}{\CheckmarkBold} & \textcolor{blue}{\CheckmarkBold} &\textcolor{red}{\XSolidBold} & 2019\\ \cline{1-1} \cline{4-8}
REACT\cite{pei2019joint} &  &  & \textcolor{red}{\XSolidBold} & \textcolor{blue}{\CheckmarkBold} & \textcolor{red}{\XSolidBold} & \textcolor{red}{\XSolidBold} & 2019\\ \cline{1-1} \cline{4-8}
SPaE\cite{shi2019unifying} &  &  & \textcolor{blue}{\CheckmarkBold} & \textcolor{blue}{\CheckmarkBold} & \textcolor{red}{\XSolidBold} & \textcolor{red}{\XSolidBold} & 2019\\ \cline{1-3} \cline{4-8}
struc2vec\cite{ribeiro2017struc2vec} & \multirow{6}{*}{\tabincell{c}{random \\ walk-based \\ methods \\ (Sec.\ref{sec:randomwalk})}}& \multirow{3}{*}{\tabincell{c}{on  similarity-biased \\ random walks (Sec.\ref{sec:randomwalk-similarity}) }} & \textcolor{blue}{\CheckmarkBold} & \textcolor{blue}{\CheckmarkBold} & \textcolor{red}{\XSolidBold} & \textcolor{red}{\XSolidBold} & 2017\\ \cline{1-1} \cline{4-8}
SPINE\cite{ijcai2019-333} &  &  & \textcolor{red}{\XSolidBold} & \textcolor{blue}{\CheckmarkBold} & \textcolor{red}{\XSolidBold} & \textcolor{red}{\XSolidBold} & 2019\\ \cline{1-1} \cline{4-8}
struc2gauss\cite{pei2020struc2gauss} &  &  & \textcolor{blue}{\CheckmarkBold} & \textcolor{blue}{\CheckmarkBold} & \textcolor{red}{\XSolidBold} & \textcolor{red}{\XSolidBold} & 2020\\ \cline{1-1} \cline{4-8} \cline{3-3}
Role2Vec\cite{ahmed2019role2vec} &  & \multirow{3}{*}{\tabincell{c}{on  feature-based \\ random walks (Sec.\ref{sec:randomwalk-feature}) }} & \textcolor{red}{\XSolidBold}  & \textcolor{red}{\XSolidBold} & \textcolor{red}{\XSolidBold} &\textcolor{blue}{\CheckmarkBold} & 2019\\ \cline{1-1} \cline{4-8}
RiWalk\cite{xuewei2019riwalk} &  &  & \textcolor{red}{\XSolidBold} & \textcolor{blue}{\CheckmarkBold} & \textcolor{red}{\XSolidBold} & \textcolor{red}{\XSolidBold} & 2019\\ \cline{1-1} \cline{4-8}
NODE2BITS\cite{jin2019node2bits} &  &  & \textcolor{red}{\XSolidBold} &  \textcolor{red}{\XSolidBold} & \textcolor{blue}{\CheckmarkBold} & \textcolor{red}{\XSolidBold} & 2019\\ \cline{1-3} \cline{4-8}
DRNE\cite{tu2018deep} & \multirow{5}{*}{\tabincell{c}{deep \\ learning \\ (Sec.\ref{sec:deeplearning})}}& \multirow{5}{*}{\tabincell{c}{via structural information \\ reconstruction/guidance \\ (Sec.\ref{sec:structure-guidance}) }} & \textcolor{blue}{\CheckmarkBold} & \textcolor{blue}{\CheckmarkBold} & \textcolor{red}{\XSolidBold} & \textcolor{red}{\XSolidBold} & 2018\\ \cline{1-1} \cline{4-8}
GAS\cite{guo2020role} &  &  & \textcolor{blue}{\CheckmarkBold} & \textcolor{blue}{\CheckmarkBold} & \textcolor{red}{\XSolidBold} & \textcolor{red}{\XSolidBold} & 2020\\\cline{1-1} \cline{4-8}
RESD\cite{zhang2021role} &  &  & \textcolor{blue}{\CheckmarkBold} & \textcolor{blue}{\CheckmarkBold} & \textcolor{blue}{\CheckmarkBold} & \textcolor{red}{\XSolidBold} & 2021\\ \cline{1-1} \cline{4-8}
GraLSP\cite{jin2020gralsp} &  &  & \textcolor{blue}{\CheckmarkBold} & \textcolor{blue}{\CheckmarkBold} & \textcolor{red}{\XSolidBold} & \textcolor{blue}{\CheckmarkBold} & 2020\\\cline{1-1} \cline{4-8}
GCC\cite{qiu2020gcc} &  &  & \textcolor{red}{\XSolidBold} & \textcolor{blue}{\CheckmarkBold} & \textcolor{blue}{\CheckmarkBold} & \textcolor{red}{\XSolidBold} & 2020\\ \cline{1-3} \cline{4-8} \cline{3-3}

\end{tabular}
\end{table*}


\subsection{Low-rank Matrix Factorization}
\label{sec:lowrank}

Low-rank matrix factorization is the most commonly used method for role-oriented embeddding methods. They generate embeddings by factorizing  matrices preserving the role similarities between nodes implicitly (i.e. feature matrices) or explicitly (i.e. similarity matrices).

\subsubsection{Structural Feature Matrix Factorization}
\label{sec:lowrank-feature}

\noindent\textbf{RolX}~\cite{henderson2012rolx}. RolX takes the advantages of feature extraction method ReFeX~\cite{henderson2011s} by decomposing the ReFeX feature matrix $\mathbf{F}_{ReFeX} \in \mathbb{R}^{n \times f}$. ReFeX firstly computes some primary features such as degree and clustering coefficient for each node. Then it aggregates neighbors' features with sum- and mean-aggregator recursively. In $k$ recursive steps, it can capture very thorough features to express the structure of $(k+1)$-hop reachable neighborhood. Non-negative Matrix Factorization (NMF) is used for generating embeddings as it is efficient compared with other matrix decomposition methods. The non-negative constraints are adapted to interpretation of roles. Thus, RolX aims to obtain two low-rank matrices as follows:

\begin{equation}
\label{eq:rolx1}
    \min_{\mathbf{H},\mathbf{M}}  \Bigl| \Bigl| \mathbf{F}_{ReFeX}-\mathbf{HM} \Bigr| \Bigr|_{F}^2, \ s.t. \ \mathbf{H,M} \ge 0 
\end{equation}
where $\mathbf{H} \in \mathbb{R}^{n \times r}$ is the embedding matrix (or role assignment matrix) and the matrix $\mathbf{M} \in \mathbb{R}^{r \times f}$ (role definition matrix) describes the contributions of each role to structural features. $r$ is the number of hidden roles which is determined by Minimum Description Length (MDL)~\cite{rissanen1978modeling}.

\noindent\textbf{GLRD}~\cite{gilpin2013guided}. GLRD extends RolX by adding different optional constraints to objective function (\ref{eq:rolx1}). 
Sparsity constraint ($\forall i, \left \| \mathbf{H}_{\cdot i} \right \|_{1} \le \epsilon_\mathbf{H} \land \left \| \mathbf{M}_{i \cdot} \right \|_{1} \le \epsilon_\mathbf{M}$) is defined for more definitive role assignments and definitions while diversity constraint ($\forall i \ne j , \mathbf{H}_{\cdot i}^\top \mathbf{H}_{\cdot j} \le \epsilon_\mathbf{H} \land \mathbf{M}_{i \cdot}\mathbf{M}_{j \cdot}^\top \le \epsilon_\mathbf{M}$) is for reducing the overlapping. $\mathbf{H}^*$ and $\mathbf{M}^*$ are previously discovered role assignments and definitions with which alternativeness constraint ($\forall i \ne j , \mathbf{H}_{\cdot i}^{*\top}\mathbf{H}_{\cdot j} \le \epsilon_\mathbf{H} \land \mathbf{M}_{i \cdot}^*\mathbf{M}_{j \cdot}^\top \le \epsilon_\mathbf{M}$) can be used for mining roles unknown.

\noindent\textbf{RID$\large\boldsymbol{\varepsilon}$Rs}~\cite{gupte2017role}. 
RID$\large{\varepsilon}$Rs uses $\varepsilon$-equitable refinement ($\varepsilon$ER) to partition nodes into different cells and compute graph-based features. An $\varepsilon$-equitable refinement partition $\pi=\{\mathcal{C}_1,\mathcal{C}_2,...,\mathcal{C}_K\}$ of  $\mathcal{V}$ satisfies the following 
rule:
\begin{equation}
\label{eq:riders1}
|\mathrm{deg}(u,\mathcal{C}_j)-\mathrm{deg}(v,\mathcal{C}_j)| \le \varepsilon, \forall u,v \in \mathcal{C}_j, \forall 1 \le i,j \le K
\end{equation}
where $\mathrm{deg}(u,\mathcal{C}_j) = |\{ u | (u,v) \in E \land v \in \mathcal{C}_j \}|$ denotes the number of nodes in cell $\mathcal{C}_j$ connected to node $v_i$. As nodes in the same cell have similar number of connections to the nodes in another cell, $\varepsilon$ERs could capture some connectivity patterns. 

Based on the cells partitioned by $\varepsilon$ERs with an relaxation parameter $\varepsilon$, the feature matrix  is defined as $(\mathbf{F}_{\varepsilon ER}^{\varepsilon})_{ij} = |\mathcal{N}_i \cap \mathcal{C}_j|$. After prunning and binning process, the feature matrices for all $1 \le \varepsilon \le \lfloor d_{avg} \rfloor$ are concatenated as the final feature matrix $\mathbf{F}_{\varepsilon ER}$. Finally, like RolX and GLRD, NMF is applied for embedding generation while right sparsity constraint (on $\mathbf{M}$) is optional for more definitive role representations.

\noindent\textbf{GraphWave}~\cite{donnat2018learning}. GraphWave treats graph diffusion kernels as probability distributions over networks and gets embeddings by using characteristic functions of the distributions. Specifically, take the heat kernel $g_\varsigma(\lambda) = e^{-\lambda \varsigma}$ with scaling parameter $\varsigma$ as an example, the spectral graph wavelets $\boldsymbol{\Psi} \in \mathbb{R}^{n \times n}$ are defined as:
\begin{equation}
\label{eq:graphwave1}
        \boldsymbol{\Psi} = \boldsymbol{\mathcal{I}}\mathbf{U} \mathrm{Diag}(g_\varsigma(\lambda_1),...,g_\varsigma(\lambda_n)) \mathbf{U}^\top 
\end{equation}
where $\boldsymbol{\mathcal{I}}$ is the one-hot encoding matrix on $\mathcal{V}$ and the scaling parameter $\varsigma$ is omitted. The $i$-th row $\boldsymbol{\Psi}_i$ represents the resulting signal from a Dirac signal around node $v_i$. Considering the empirical characteristic function:
\begin{equation}
\label{eq:graphwave2}
\varphi_i(t) = \frac{1}{n} \sum_{j=1}^{n} e^{\mathfrak{i}t\boldsymbol{\Psi}_{ij}}
\end{equation}
where $\mathfrak{i}$ denotes the imaginary number, $v_i$'s embedding vector $\mathbf{H}_i$ is generated by concatenating pairs of  $\mathrm{Re}(\varphi_i(t))$ and $\mathrm{Im}(\varphi_i(t))$ at $r$ evenly spaced points $t_1,...,t_r$.

\noindent\textbf{HONE}~\cite{rossi2020structural}. HONE constructs weighted motif graphs in which the weight of an edge is the count of the co-occurrences of the two endpoints in a specific motif. For a motif represented by its weighted motif adjacency matrix $ \mathbf{A}_{\mathcal{M}_m}$, HONE characters the higher-order structure by deriving matrices from its k-step matrices $\mathbf{A}_{\mathcal{M}_m}^k$. These new matrices are designed by imitating some popular matrices based on normal adjacency matrix such as transition matrix $\mathbf{P} = \mathbf{D}^{-1}\mathbf{A}$ and Laplacian matrix $\mathbf{L} = \mathbf{D}-\mathbf{A}$. Here we use $(\mathbf{F}_{HONE})_{\mathcal{M}_m}^{(k)}$ to denote the derived matrices. Then the k-step embeddings can be learned as:
\begin{equation}
    \mathop{\arg\min}_{\mathbf{H}_{\mathcal{M}_m}^{(k)},\mathbf{M}_{\mathcal{M}_m}^{(k)}} \mathbb{D}_{Breg}(\mathbf{F}^{(k)}_m | \Psi(\mathbf{H}_{\mathcal{M}_m}^{(k)}\mathbf{M}_{\mathcal{M}_m}^{(k)}))
    \label{eq:hone1}
\end{equation}
where $\mathbb{D}_{Breg}$ is the Bregman divergence and $\Psi(\cdot)$ is a matching function. The global embeddings are generated by minimizing the following objective:
\begin{equation}
    \min_{\mathbf{H},\mathbf{M}} \Bigl| \Bigl| \mathbf{F}_{HONE}-\mathbf{HM} \Bigr| \Bigr|_{F}^2
    \label{eq:hone2}
\end{equation}
where $\mathbf{F}_{HONE}$ is obtained by concatenating the $\mathbf{H}_{\mathcal{M}_m}^{(k)}$ with all the considered motifs and steps. If necessary, attributes diffused by transition matrix based on different motifs and steps can be added into $\mathbf{F}_{HONE}$.

\noindent\textbf{\textit{Remark.}} Aforementioned methods assume that nodes in similar roles have similar structural features. Thus, they apply matrix factorization on the feature matrices to obtain role-based representations. RolX, GLRD and RID$\large{\varepsilon}$Rs directly get embeddings which give soft role assignment by factorizing feature matrices. As the feature matrices are usually lower dimension, these methods are quite efficient. GraphWave uses 
eigen-decomposition and empirical characteristic function to characterize the structural patterns of each node, which leads to robust embeddings but high computation cost. The weighted motif adjacency matrices in HONE capture higher-order proximities actually, while they can obtain structural information because each matrix represents one motif.

\subsubsection{Structural Similarity Matrix Factorizaiton}
\label{sec:lowrank-similarity}

\noindent\textbf{xNetMF}~\cite{heimann2018regal}.
xNetMF is an embedding method designed for an embedding-based network alignment approach REGAL. It firstly obtains a node-to-node similarity matrix $\mathbf{S}_{REGAL}$ based on both structures and attributes:
\begin{equation}
\label{eq:xnetmf1}
\mathbf{S}_{ij} = \mathrm{exp}(- \gamma_{s} \mathrm{dist}_s(v_i,v_j) - \gamma_{a} \mathrm{dist}_a(v_i,v_j))
\end{equation}
where $\mathrm{dist}_s(v_i,v_j)$ and $\mathrm{dist}_a(v_i,v_j)$ are structure-based and attribute-based distance between node $v_i$ and node $v_j$ while $\gamma_s$ and $\gamma_a$ are balance parameters of the two distances. $\mathrm{dist}_s(v_i,v_j)$ is the Euclidean distance between node features. And $\mathrm{dist}_a(v_i,v_j)$ counts different attributes between nodes, i.e., $\mathrm{dist}_a(v_i,v_j) = |\{a | \mathbf{X}_{ia} \neq \mathbf{X}_{ja} \land 1\le a \le x\}|$. The feature matrix $\mathbf{F}_{REGAL}$ is defined by counting nodes with the same logarithmically binned degree in each node's $k$-hop reachable neighborhood as follows:
\begin{equation}
\label{eq:xnetmf2}
\begin{aligned}
\mathbf{F}_{ic}^k & = | \mathcal{D}_{i,c}^k|  =  | \{ v_j \in \mathcal{N}^k_i| \lfloor \mathrm{log}_2 d_j\rfloor= c \}|,\\
\mathbf{F}_{i} & = \sum_{k=1}^{K} \delta^k \mathbf{F}_{i}^k
\end{aligned}
\end{equation}
where $\delta^k \in (0,1]$ is a discount factor for lessening the importance of higher-hop neighbors.

Then on computed $\mathbf{S}$, matrix factorization methods can be applied for obtaining embedding matrix $\mathbf{H}$ satisfying $\mathbf{S} \approx \mathbf{H}\mathbf{M}^\top$. As the high dimension and rank of $\mathbf{S}$ lead to high computation, an implicit matrix factorization approach extending Nyström method~\cite{drineas2005nystrom} is proposed as follows:
\begin{enumerate}
    \item Select $r\ll n$ nodes as landmarks randomly or based on node centralities.
    \item Compute a node-to-landmark similarity matrix $\mathbf{C} \in \mathbb{R}^{n \times r}$ with Eq.(\ref{eq:xnetmf1}) and extract a landmark-to-landmark similarity matrix $\mathbf{B} \in \mathbb{R}^{r \times r}$ from $\mathbf{C}$.
    \item Apply Singular Value Decomposition on the pseudoinverse of $\mathbf{B}$ so that $\mathbf{B}^\dagger = \mathbf{V}\mathbf{\Sigma}\mathbf{Y}^\top$.
    \item Obtain embedding matrix $\mathbf{H}$ by computing and normailize $\mathbf{CV\Sigma}^{- \frac{1}{2}}$.
    
\end{enumerate}
With above method, embeddings are actually generated by factorizing a low-rank approximation of $\mathbf{S}$, i.e., $\widetilde{\mathbf{S}} = \mathbf{C(V \Sigma Y}^\top)\mathbf{C}^\top$.
Meanwhile, the computation can be reduced, as only a small matrix $\mathbf{B}^\dagger$ is decomposed. 

\noindent\textbf{EMBER}~\cite{jin2019smart}. EMBER is designed for mining professional roles in weighted directed email networks. It defines node outgoing feature matrix $\mathbf{F}_{EMBER}^+$ as:
\begin{equation}
\label{eq:ember1}
\begin{aligned}
\mathbf{F}_{ic}^{k+} &= \sum_{v_j \in \mathcal{D}^{k+}_{i,c}} \mathrm{pw}(\mathcal{P}_{v_i \rightarrow v_j}^{k+}),\\
\mathbf{F}_{i}^{+} &= \sum_{k=1}^{K} \delta^k \mathbf{F}_{i}^{k+}
\end{aligned}
\end{equation}
where $\mathrm{pw}(\mathcal{P}_{v_i \rightarrow v_j}^{k+})$ denotes the product of all edge weights in a $k$-step shortest outgoing path $\mathcal{P}_{v_i \rightarrow v_j}^{k+}$. The incoming feature matrix $\mathbf{F}$ is defined similarly. By concatenating the incoming and outgoing feature matrices, the final feature matrix $\mathbf{F}_{EMBER} = [\mathbf{F}^+,\mathbf{F}^-]$ is obtained. The node-to-node similarities are computed through Eq.(\ref{eq:xnetmf1}) without attribute-based distance, i.e., $\mathbf{S}_{ij} = \mathrm{exp}(-\left \| \mathbf{F}_i -  \mathbf{F}_j \right \|^2)$. EMBER uses the same implicit matrix factorization approach to generate embeddings. Note that if the feature extraction part of EMBER is applied on an undirected unweight network, EMBER 
will be equivalent to xNetMF without attributes.

\noindent\textbf{SEGK}~\cite{nikolentzos2019learning}. SEGK leverages graph kernels to compute node structural similarities. To compare the structure more carefully, it computes node similarities with different scales of neighborhood as follows:
\begin{equation}
\label{eq:segk1}
\mathbf{S}_{ij} =\sum_{k=1}^K \hat{\mathcal{K}}(\mathcal{G}_i^k,\mathcal{G}_j^k) \hat{\mathcal{K}}(\mathcal{G}_i^{k-1},\mathcal{G}_j^{k-1})
\end{equation}
where $\hat{\mathcal{K}}(\mathcal{G}_i^{0},\mathcal{G}_j^{0}) = 1$ and $\hat{\mathcal{K}}$ denotes the normalized kernel which is defined as:
\begin{equation}
\label{eq:segk2}
\hat{\mathcal{K}}(\mathcal{G},\mathcal{G}') = \frac{\mathcal{K}(\mathcal{G},\mathcal{G}')}{\sqrt{\mathcal{K}(\mathcal{G},\mathcal{G}) \mathcal{K}(\mathcal{G}',\mathcal{G}')}}
\end{equation}
SEGK chooses the shortest path kernel, Weisfeiler-Lehman subtree kernel, or graphlet kernel for practical use of $\mathcal{K}(\cdot,\cdot)$. Then Nyström method~\cite{williams2001using} is employed on the factorization of $\mathbf{S}$ for efficient computation and low dimensions of embeddings as follows:
\begin{equation}
\label{eq:segk3}
\mathbf{H}=\mathbf{S}\mathbf{U}_{[r]}\mathbf{\Lambda}^{-\frac{1}{2}}_{[r]}
\end{equation}
where $\mathbf{U}_{[r]}$ denotes the matrix of first $r$ eigenvectors and $\mathbf{\Lambda}_{[r]}$ is the diagonal matrix of corresponding eigenvalues.

\noindent\textbf{REACT}~\cite{pei2019joint}. REACT aims to detect communities and discover roles by applying non-negative matrix tri-factorization on RoleSim~\cite{jin2011axiomatic} matrix $\mathbf{S}$ and adjacency matrix $\mathbf{A}$, respectively. RoleSim matrix is developed with the idea of regular equivalence and is a pair-wise similarity matrix computed by iteratively updating the following scores:

\begin{equation}
    \mathbf{S}_{ij}=(1-\beta) \max _{\mathrm{M}(v_i, v_j)} \frac{\sum_{(v_{i'}, v_{j'}) \in \mathrm{M}(v_{i}, v_{j})} \mathbf{S}_{i'j'}}{d_{i}+d_{j}-|\mathrm{M}(v_i, v_j)|}+\beta
    \label{eq:rolesim}
\end{equation}
where $\mathrm{M}(v_i, v_j)$ is a matching between the neighborhoods of $v_i$ and $v_j$, and $\beta$ ($0 < \beta <1$) is a decay factor. In addition, $L_{2,1}$ norm is leveraged as the regularization to make the distribution of roles within communities as diverse as possible.
Thus, the objective function of REACT is:
\begin{equation}
\label{eq:react1}
\begin{aligned}
    \quad &\min_{\mathbf{H}_R,\mathbf{M}_R,\mathbf{H}_C,\mathbf{M}_C}  \  \left \| \mathbf{S}-\mathbf{H}_R\mathbf{M}_R\mathbf{H}_R^\top \right \|_{F}^2 \\
    \quad &  \quad + \left \| \mathbf{A}-\mathbf{H}_C\mathbf{M}_C\mathbf{H}_C^\top \right \|_{F}^2  +  \gamma_{2,1} \left \| \mathbf{H}_C^\top\mathbf{H}_R \right \|^{\qquad}_{2,1}, \\
    s.t. \quad &\mathbf{H}_R,\mathbf{M}_R,\mathbf{H}_C,\mathbf{M}_C \ge 0, \mathbf{H}_R^\top \mathbf{H}_R = \mathbf{I},\mathbf{H}_C^\top \mathbf{H}_C = \mathbf{I}.
\end{aligned}
\end{equation}
where $\mathbf{H}_R$/$\mathbf{H}_C$ denotes the embedding matrix for roles/communities, and $\mathbf{M}_R$/$\mathbf{M}_C$ denotes the interaction between roles/communities. $\gamma_{2,1}$ is the weight of regularization. Orthogonal constraint on embedding matrices is added for increased interpretability.

\noindent\textbf{SPaE}~\cite{pei2019joint}. SPaE also tries to capture communities and roles simultaneously. For node structural similarity, it computes cosine similarity between the standardized Graphlet Degree Vectors of nodes, and generates role-based embeddings via Laplacian eigenmaps method as follows:
\begin{equation}
\label{eq:spae1}
    \max_{\mathbf{H}_R} \mathcal{J}_R = \mathrm{Tr}(\mathbf{H}_R^\top \mathbf{L_S} \mathbf{H}_R), \ 
    s.t. \  \mathbf{H}_R^\top \mathbf{H}_R = \mathbf{I}.
\end{equation}
where $\mathbf{L_S}$ is the symmetric normalized matrix of structural similarity matrix $\mathbf{S}_{SPaE}$. SPaE obtains community-based embeddings similarly as follows:
\begin{equation}
\label{eq:spae2}
    \max_{\mathbf{H}_C} \mathcal{J}_C = \mathrm{Tr}(\mathbf{H}_C^\top \mathbf{L_\mathbf{A}} \mathbf{H}_C), \ 
    s.t. \  \mathbf{H}_C^\top \mathbf{H}_C = \mathbf{I}.
\end{equation}
where $\mathbf{L_\mathbf{A}} = \mathbf{D}^{-\frac{1}{2}}\mathbf{A}\mathbf{D}^{-\frac{1}{2}}$ is the symmetric normalized adjacency matrix. To map $\mathbf{H}_R$ and $\mathbf{H}_C$ into a unified embedding space, SPaE generates hybrid embeddings by maximizing the following objective function:
\begin{equation}
\begin{aligned}
\label{eq:spae3}
    &\max_{\mathbf{H}_R,\mathbf{H}_C,\mathbf{H}_H} \ \mathcal{J}_R + \mathrm{p}_R
    + \gamma (\mathcal{J}_C + \mathrm{p}_C), \\
    &s.t. \ \mathbf{H}_R^\top \mathbf{H}_R = \mathbf{I}, \mathbf{H}_C^\top \mathbf{H}_C = \mathbf{I}, \mathbf{H}_H^\top \mathbf{H}_H = \mathbf{I}.
\end{aligned}
\end{equation}
where $\mathbf{H}_H$ denotes the hybrid embedding matrix and $\gamma$ is the balance parameter. $\mathrm{p}_R = \mathrm{Tr}(\mathbf{H}_R^\top\mathbf{H}_H\mathbf{H}_H^\top\mathbf{H}_R)$ and $\mathrm{p}_C = \mathrm{Tr}(\mathbf{H}_C^\top\mathbf{H}_H\mathbf{H}_H^\top\mathbf{H}_C)$.

\noindent\textbf{\textit{Remark.}} These methods all explicitly compute structural similarities based on features, e.g., graph kernels, role equivalence, and so on. Most of them have considered the similarities between multiple hops of neighborhoods. Their effectiveness on role discovery depends on the quality of the similarity matrices. One major problem of this kind of methods is the issue of efficiency: computing pair-wise similarity and factorizing the high-dimensional similarity matrix $\mathbf{S} \in \mathbb{R}^{n \times n}$ are time-consuming. So xNetMF, EMBER and SEGK apply Nyström method to improve the efficiency as their similarity matrices are Gram matrices~\cite{drineas2005nystrom}. 

\subsection{Shallow Models Using Random Walks}
\label{sec:randomwalk}



\begin{figure}
    \centering
    \subfigure[Normal Random Walks]{\includegraphics[width=0.8\columnwidth]{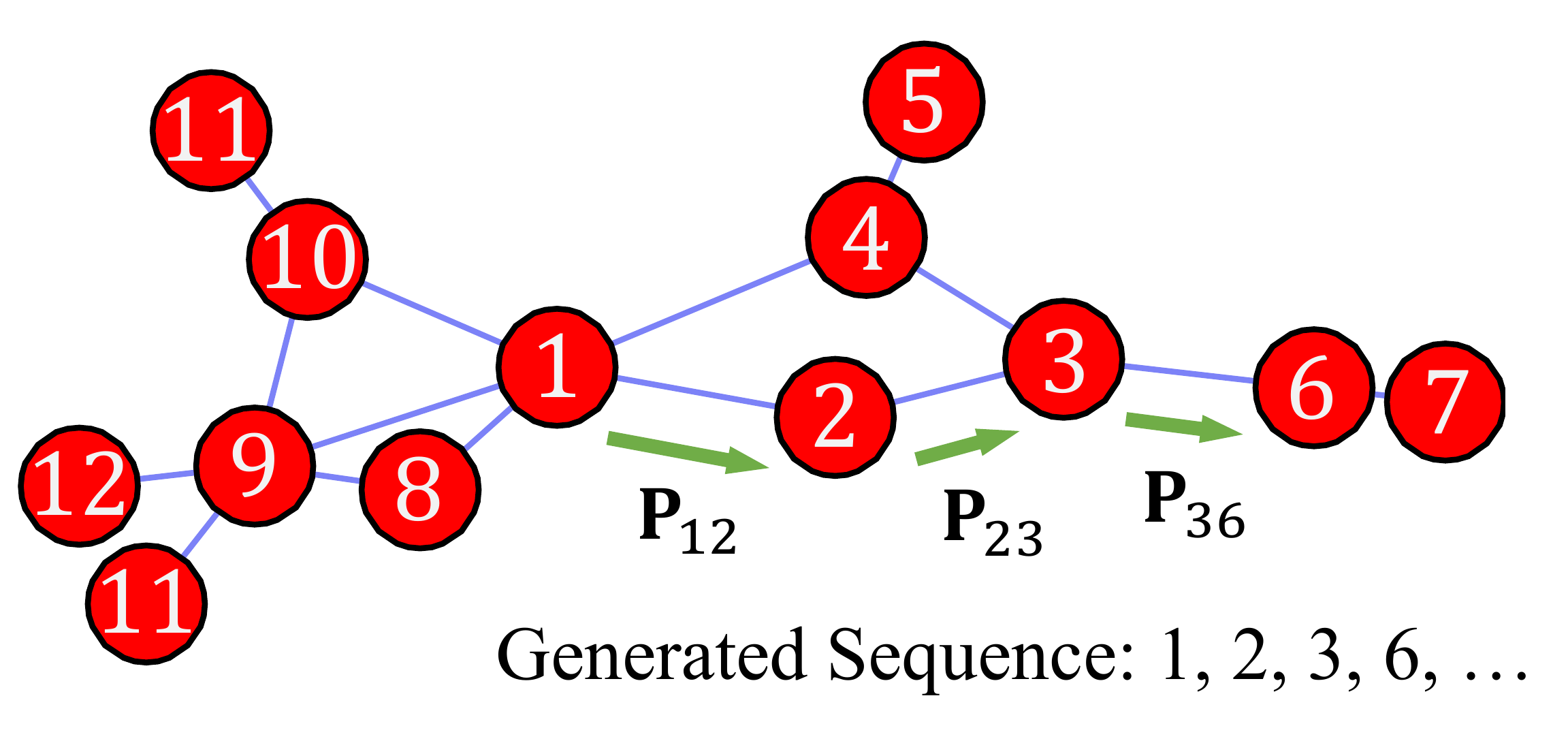}}
    \subfigure[Structural Similarity-biased Random Walks]{
\includegraphics[width=0.8\columnwidth]{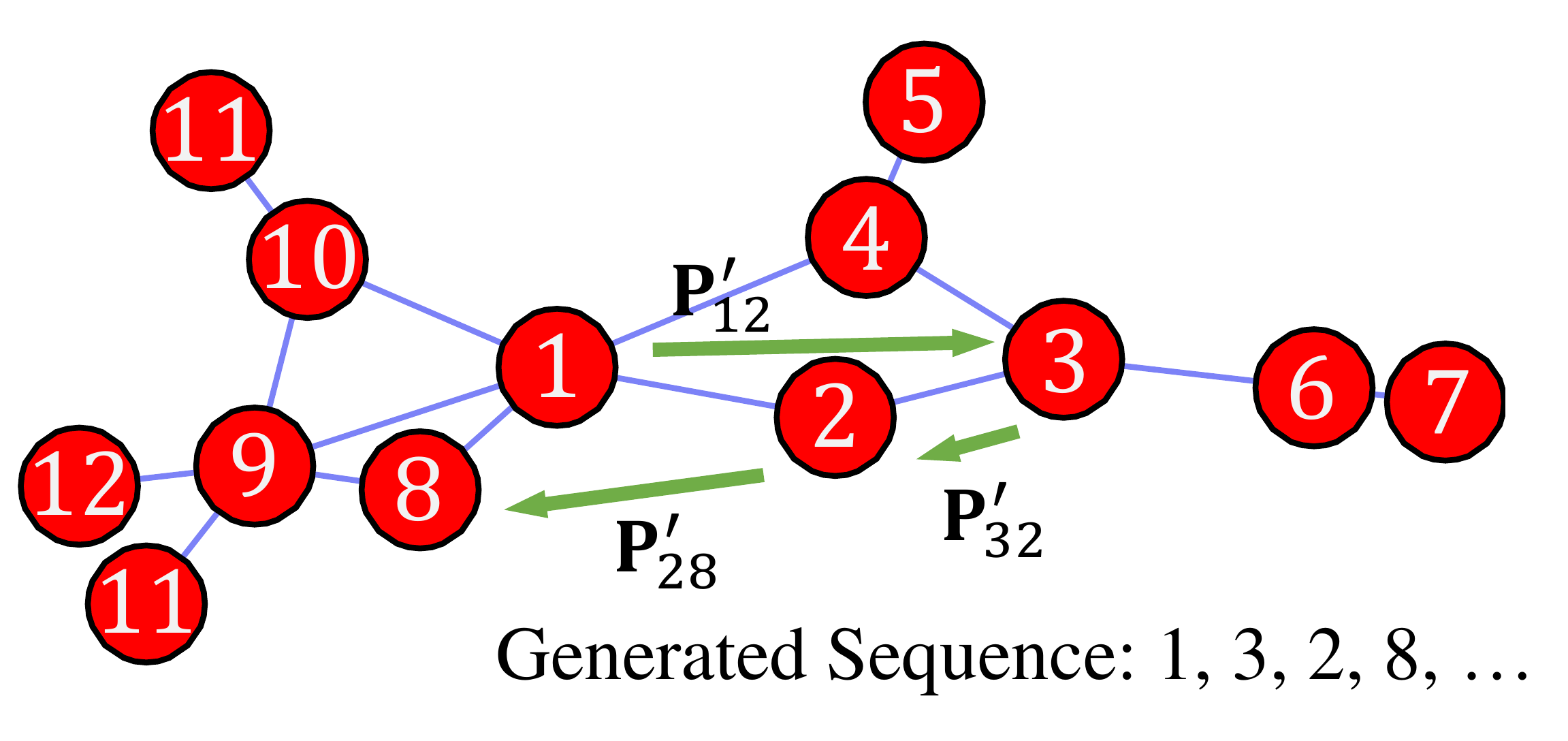}}
\subfigure[Structural Feature-based Random Walks]{
\includegraphics[width=0.8\columnwidth]{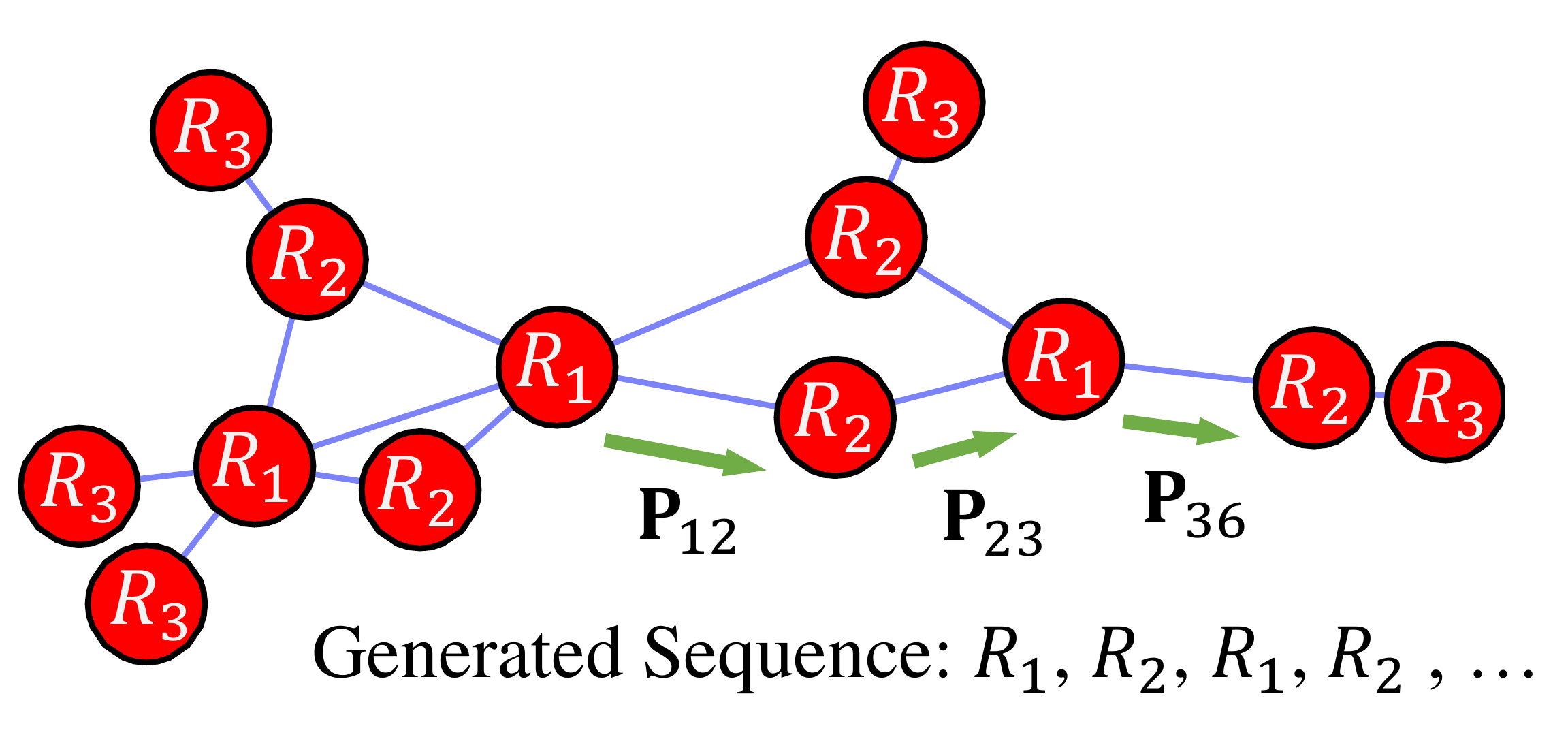}}
    \caption{Different types of random walks. Note that $\mathbf{P}'$ is the biased transition matrix computed based on node structural similarities. $R_i$ is the role indicator mapped from node structural features.}
\label{fig:randomwalks}
\end{figure}

Random walk is a common way to capture node proximity used by network embedding methods~\cite{perozzi2014deepwalk,grover2016node2vec}. 
Recently, two strategies have been proposed to adapt random walks to role-oriented tasks: (1) structural similarity-biased random walks makes structurally similar nodes more likely to appear in the same sequence  (as shown in Fig.~\ref{fig:randomwalks}(b)). (2) structural feature-based random walks, e.g., attributed random walks~\cite{ahmed2019role2vec}, map nodes with similar structural features to the same role indicator and replace ids in random walk sequences with the indicators (see Fig.~\ref{fig:randomwalks}(c)). The first way can preserve structural similarity into co-occurrence relations of nodes in the walks. While the second way preserves structural similarity into role indicators and may capture the proximity between roles through the co-occurrence relations of the indicators as well.

Usually, language models such as Skip-Gram~\cite{mikolov2013distributed} are applied on generated random walks to map the similarities into embedding vectors~\cite{perozzi2014deepwalk,grover2016node2vec,ribeiro2017struc2vec,ahmed2019role2vec}. 
However, some different mapping mechanisms are also employed such as the SimHash~\cite{charikar2002similarity} used in NODE2BITS~\cite{jin2019node2bits}.

\subsubsection{Structural Similarity-biased Random Walks}
\label{sec:randomwalk-similarity}

\noindent\textbf{struc2vec}~\cite{ribeiro2017struc2vec}. Struc2vec generates structurally biased id-based contexts via random walks on a hierarchy of constructed complete graphs. 

In detail, it firstly computes structural distances between a pair of nodes as follows:
\begin{equation}
\label{eq:struc2vec1}
\begin{aligned}
        \mathrm{dist}_d^k(v_i,v_j) = & \mathrm{dist}_d^{k-1}(v_i,v_j) +  \mathrm{DTW}(\mathcal{H}^k_i,\mathcal{H}^k_j), \\
        & 0 \le k \le k^*
\end{aligned}
\end{equation}
where $\mathrm{DTW}(\cdot,\cdot)$ denotes Dynamic Time Warping (DTW). $\mathrm{d}(a,b) = \mathrm{max}(a,b) / \mathrm{min}(a,b) -1$ is  adopted as the distance function for DTW. $k^*$ is the diameter of the $\mathcal{G}$. $\mathcal{H}^k_i$ is the ordered degree sequence of nodes at the exact distance $k$ from $v_i$. Note that $\mathcal{H}^0_i = \{ d_i \}$ and $\mathrm{dist}_d^{-1}(v_i,v_j)$ is set to the constant 0.

Then a multi-layer weighted context graph $\mathcal{G}_C = (\mathcal{V}_C,\mathcal{E}_C)$ is built. Each layer $k = 0,...,k^*$ is an undirected complete graph $\mathcal{G}_C^k
= (\mathcal{V}_C^k,\mathcal{E}_C^k)$. $\mathcal{V}_C^k = \{ v_1^k,...,v_n^k \}$ 
where the corresponding node of $v_i \in \mathcal{V}$ in $k$-layer is denoted as $v^k_i$. The weight $w_C^k(v_i^k,v_j^k)$ of edge $(v_i^k,v_j^k) \in \mathcal{E}_C^k$ is defined as follows:
\begin{equation}
\label{eq:struc2vec2}
    w_C^k(v_i^k,v_j^k) = 
\mathrm{exp}(-\mathrm{dist}_d^k(v_i^k,v_j^k)) , k = 0,...,k^*
\end{equation}
The neighboring layers are connected through directed edges between the corresponding nodes. Thus $\mathcal{V}_C = \bigcup_0^{k^*} \mathcal{V}_C^k$ and $\mathcal{E}_C = (\bigcup_0^{k^*} \mathcal{E}_C^k) \cup (\bigcup_1^{k^*} \bigcup_1^{n} \{ (v_i^k,v_i^{k-1}) \}) \cup (\bigcup_0^{k^*-1} \bigcup_1^{n} \{ (v_i^{k-1},v_i^{k}) \})$. The edge weights between layers are defined as follows:
\begin{equation}
\label{eq:struc2vec3}
\begin{aligned}
    w_C(v_i^k,v_i^{k+1}) = & 
\mathrm{log}(\Gamma(v_i^k) + \mathrm{e}), k = 0,...,k^*-1 \\
 w_C(v_i^k,v_i^{k-1}) = & 
1, k = 1,...,k^*
\end{aligned}
\end{equation}
where $\Gamma(v_i^k)$ counts the edge $(v_i^k,v_j^k) \in \mathcal{E}_C^k$ whose weight is larger than the average edge weight of $\mathcal{G}_C^k$. That is:
\begin{equation}
\label{eq:struc2vec4}
\begin{aligned}
\Gamma(v_i^k) = |\{ (v_i^k,v_j^k) | w_C^k(v_i^k,v_j^k) > \\ \frac{\sum_{(v^k_{i'},v^k_{j'})\in \mathcal{E}_C^k} w_C^k(v^k_{i'},v^k_{j'})}{\tbinom{n}{2}} \}|
\end{aligned}
\end{equation}

Then id-based random walks can be employed on $\mathcal{G}_C$ and started in layer $0$ for context generation of each node. In detail, the walk stays in the current layer with a given probability $q_b$. In this situation, the probability of a walk from $v_i^k$ to $v_j^k$ is:
\begin{equation}
\label{eq:struc2vec5}
(\mathbf{P}^k_{S2V})_{ij} = \frac{w_C^k(v_i^k,v_j^k))}{\sum_{(v_i^k,v_{j'}^k) \in \mathcal{E}^k_C}w_C^k(v_i^k,v_{j'}^k))}
\end{equation}
With probability $1-q_b$, the walk steps across layers with the following stepping probability:
\begin{equation}
\label{eq:struc2vec6}
\begin{aligned}
  p(v_i^k,v_i^{k+1}) = & \frac{w_C(v_i^k,v_i^{k+1})}{w_C(v_i^k,v_i^{k+1})+w_C(v_i^k,v_i^{k-1})}  \\
  p(v_i^k,v_i^{k-1}) = & 1 - p(v_i^k,v_i^{k+1})
\end{aligned}
\end{equation}
Note that $v_i^k$ with different $k$ have the same id in the context.

On the structural context, struc2vec leverages Skip-Gram with Hierarchical Softmax to learn embeddings.

\noindent\textbf{SPINE}~\cite{ijcai2019-333}. SPINE uses largest $f$ values of $i$th row of Rooted PageRank Matrix $\boldsymbol{\Omega} = (1 - \beta_{RPR})(\mathbf{I} - \beta_{RPR}\mathbf{P})^{-1}$ as $v_i$'s feature $(\mathbf{F}_{RPR})_i$. $\beta_{RPR}$ is the probability of stepping to a neighbor, while with probability $1-\beta_{RPR}$, a walk steps back to the start node. For the inductive setting, SPINE computes $\mathbf{F}_i$ via a Monte Carlo approximation. 

To simultaneously capture structural similarity and proximity, SPINE designs a biased random walk method. With probability $\varrho_{br}$, the walk steps to a structural similar node based on the following transition matrix:
\begin{equation}
\label{eq:spine1}
(\mathbf{P}^k_{SPINE})_{ij} = \frac{\mathrm{sim}(v_i,v_j)}{\sum_{v_k \in \mathcal{V},v_k \ne v_i} \mathrm{sim}(v_i,v_k)}
\end{equation}
Here $\mathrm{sim}(\cdot,\cdot)$ can be computed via DTW or other methods based on node features. With probability $1-\varrho_{br}$, normal random walks are applied. Thus, with larger $\varrho_{br}$, SPINE can be more role-oriented. The embeddings are learned through Skip-Gram with Negative Sampling (SGNS). To leverage attributes, the embeddings are generated as:
\begin{equation}
\label{eq:spine2}
\mathbf{H}_{i} = \sigma(\sum_{j=1}^{f} \mathbf{F}_{ij}\mathbf{X}^{<i,f>}_j\mathbf{W})
\end{equation}
where $\mathbf{X}^{<i,f>}$ represents the attribute matrix of which the rows correspond to $f$ largest values of $\boldsymbol{\Omega}_{i}$. $\mathbf{W}$ is the weight matrix of the multi-layer perceptron (MLP). 
\noindent\textbf{struc2gauss}~\cite{pei2020struc2gauss}. For each node $v_i$, struc2gauss generates a Gaussian distribution: $\mathcal{Z}_i = \mathrm{Gauss}(\boldsymbol{\mu}_i, \boldsymbol{\Sigma}_i)$ 
to model both structural similarity and uncertainty. After calculating structural similarity via existing methods such as RoleSim~\cite{jin2011axiomatic}, it samples the top-$K$ most similar nodes for a node as its positive set $\mathcal{S}^+_i$. The positive sampling of struc2gauss could be regarded as special random walks with mandatory restart on a star-shaped graph where the star center is the target node and star edges are the most similar nodes. The negative sample set $\mathcal{S}^-_i$ has the same size of $\mathcal{S}^+_i$ and is generated as in the normal random-walk based methods. To push the Gaussian embeddings of similar nodes closer and those of dissimilar nodes farther, struc2gauss uses the following max-margin ranking objective:
\begin{equation}
\label{eq:struc2gauss1}
\mathcal{L}=\sum_{v_i \in \mathcal{V}} \sum_{v_j \in \mathcal{S}^+_i} \sum_{v_k \in \mathcal{S}^-_i} \max \left(0, m-\mathrm{sim}\left(\mathcal{Z}_{i}, \mathcal{Z}_{j}\right)+\mathrm{sim}\left(\mathcal{Z}_{i}, \mathcal{Z}_{k}\right)\right)
\end{equation}
where $m$ is the margin parameter to push dissimilar distributions apart, and $\mathrm{sim}\left(\mathcal{Z}_{i}, \mathcal{Z}_{j}\right)$ is the similarity measure between distributions of $v_i$ and $v_j$. There are different similarity measures that can be used such as logarithmic inner product and KL divergence. For normal tasks, the mean vectors of those Gaussian distributions can be treated as embeddings, i.e., $\mathbf{H}_i = \boldsymbol{\mu}_i$.

\noindent\textbf{\textit{Remark.}} These methods reconstruct the edges between nodes based on the structural similarities so that the context nodes obtained by random walks are structurally similar to the central nodes. Compared with SPINE and struc2gauss, struc2vec clearly construct edges that better represent role information in the multi-layer complete graphs, which leads to better embeddings but higher time and space complexities.

\subsubsection{Structural Feature-based Random Walks} 
\label{sec:randomwalk-feature}

\noindent\textbf{Role2Vec}~\cite{ahmed2019role2vec}. Role2Vec firstly maps nodes into several disjoint roles. Logarithmically binning, K-means with low-rank factorization and other methods on features and attributes can be chosen for the role mapping $\phi: \mathcal{V} \rightarrow \mathcal{R}$. Motif-based features, such as Graphlet Degree Vectors, are recommended since motif can better capture the high-order structural information. 

Then random walks are performed but the ids in generated sequences are replaced with role indicators. With feature-based role context, the language model CBOW can be used for obtaining embeddings of roles. Nodes partitioned into the same role have the same embeddings 

\noindent\textbf{RiWalk}~\cite{xuewei2019riwalk}. RiWalk designs structural node indicators approximating graph kernels. In a given subgraph $\mathcal{G}_i^k$, the indicator approximating shortest path kernel for node $v_j \in \mathcal{N}_i^k$ is defined as the concatenation of the degrees of $v_i$ and $v_j$ and the shortest path length between them:
\begin{equation}
\label{eq:riwalk1}
\phi_{SP}^i(v_j) = \mathrm{b}(d_i) \circ \mathrm{b}(d_j) \circ s_{ij}, v_j \in \mathcal{N}_i^k
\end{equation}
where $\mathrm{b}(x) = \lfloor \mathrm{log}(x) + 1 \rfloor$ is a logarithmically binning function.
The indicator approximating Weisfeiler-Lehman sub-tree kernel is defined as:
\begin{equation}
\label{eq:riwalk2}
\phi_{WL}^i(v_j) = \mathrm{b}(\mathbf{l}^{<i,i>}) \circ \mathrm{b}(\mathbf{l}^{<i,j>}) \circ s_{ij}, v_j \in \mathcal{N}_i^k
\end{equation}
where $\mathbf{l}^{<i,j>}$ is a vector of length $k+1$ whose $l$-th element is the count of $v_j$'s neighbers at distance $l$ to $v_i$ in $\mathcal{G}_i^k$, i.e.:
\begin{equation}
\label{eq:riwalk3}
\mathbf{l}^{<i,j>}_l = |\{ v_{j'} \in \mathcal{N}_j | s_{ij'} = l \}|, l=0,...,k
\end{equation}
Then the random walks starting from $v_i$ are performed on each $\mathcal{G}_i^k$. The nodes are relabeled indicated by Eq.(\ref{eq:riwalk1}) or Eq.(\ref{eq:riwalk2}) while only $v_i$ is not relabeled. And embeddings are learned via SGNS on the generated sequences.

\noindent\textbf{NODE2BITS}~\cite{jin2019node2bits}. NODE2BITS is designed for entity resolution on temporal networks. Here we use $\tau_{ij}$ to denote the the timestamp of edge $e_{ij}$. To integrate temporal information, NODE2BITS utilizes temporal random walks in which edges are sampled with non-decreasing timestamps. The following stepping probability is defined to capture short-term transitions in temporal walks:
\begin{equation}
\label{eq:node2bits1}
p_s(v_i,v_j) = \frac{\mathrm{exp}(-\tau_{ij}/T)}{\sum_{(v_i,v_{j'})\in \mathcal{E}} \mathrm{exp}(-\tau_{ij'}/T)}
\end{equation}
where $T$ is the maximal duration between all timestamps. The stepping probability in long-term policy is defined similarly with positive signs. Multiple walks are generated for each edge and the temporal context of different hops $\Delta t$ for a node can be extracted from the walks. Then structual features and attributes are fused in temporal walks. For each node $v_i$ with a specific $\Delta t$, histograms are applied on multi-dimensional features (and node types if the network is heterogeneous) to aggregate information in the neighborhood and they are concatenated as a vector $(\mathbf{H}_{HIST})_i^{\Delta t}$.   SimHash~\cite{charikar2002similarity} is applied by projecting the histogram $(\mathbf{H}_{HIST})_i^{\Delta t}$ to several random hyperplanes for generating binary hashcode $(H_{HASH})_i^{\Delta t}$. The final embeddings are obtained via concatenation on $(H_{HASH})_i^{\Delta t}$ across different $\Delta t$s. 

\noindent\textbf{\textit{Remark.}} The above three methods are very different on their motivations of utilizing structural feature-based random walks. Role2vec assigns roles firstly and then employs random walks with role indicators. It essentially captures proximity between assigned roles. RiWalk relabels the walks in subgraphs to approximate graph kernels. NODE2BITS uses random walks as neighbor feature aggregators.

\subsection{Deep Learning Models}
\label{sec:deeplearning}
Recently, a few works focus on leveraging deep learning techniques to role-oriented network representation learning. Though deep learning can provide more varied and powerful mapping mechanisms, it needs to be trained with more carefully designed structural information guidance.

\subsubsection{Structural Information Reconstruction/Guidance} 
\label{sec:structure-guidance}

\noindent\textbf{DRNE}~\cite{tu2018deep}. DRNE is proposed to capture regular equivalence in networks, so it learns node embeddings in a recursive way with the following loss function:
\begin{equation}
\label{eq:drne1}
\mathcal{L}_{equiv}=\sum_{v_i\in \mathcal{V}} \left \| \mathbf{H}_i - \breve{\mathbf{H}}_i \right \|_2^2
\end{equation}
where $\breve{\mathbf{H}}_i$ is the aggregation of the neighbors' embeddings via a layer normalized Long Short-Term Memory. To make the neighbor information available for LNLSTM, for each node $v_i$, it downsamples a fixed number of neighbors with large degrees and orders them based on the degrees. Denoting their embeddings as $\{\mathbf{H}_{(1)},...,\mathbf{H}_{(T)}\}$ the aggregating process is $\breve{\mathbf{H}}_{(t)} = \mathrm{LNLSTM}(\mathbf{H}_{(t)},\breve{\mathbf{H}}_{(t-1)})$ and finally $\breve{\mathbf{H}}_i = \breve{\mathbf{H}}_{(T)}$.

Additionally, DRNE proposes a degree-guided regularizer to avoid the trivial solution where all embeddings are $\mathbf{0}$. The regularizer is as follows:
\begin{equation}
\label{eq:drne2}
\mathcal{L}_{deg}=\sum_{v_i\in \mathcal{V}} (\mathrm{log}(d_i+1) - \mathrm{MLP}_{deg}(\breve{\mathbf{H}}_i))^2
\end{equation}
The regularizer with a parameter $\gamma_{deg}$ is weighed and the whole model is trained via the combined loss:
\begin{equation}
\label{eq:drne3}
\mathcal{L}=\mathcal{L}_{equiv}+\gamma_{deg}\mathcal{L}_{deg}
\end{equation}

\noindent\textbf{GAS}~\cite{guo2020role}. Graph Neural Networks have the power to capture structure as they are closely related to Weisfeiler-Lehman (WL) test in some ways~\cite{xu2018how}. GAS applies a $L$-layer graph convolutional encoder, in which each layer is :
\begin{equation}
\label{eq:gas1}
\begin{aligned}
\mathbf{H}^{(l)} = \sigma(\Tilde{\mathbf{A}}\mathbf{H}^{(l-1)}\Theta^{(l-1)})
\end{aligned}
\end{equation}
where $\Tilde{\mathbf{A}} = \mathbf{A} + \mathbf{I}$ and $\Theta^{(l-1)}$ is the parameter matrix in the $l$-th layer. The input $\mathbf{H}^{(0)}$ could be $\Tilde{\mathbf{A}}$ or an embedding lookup table. Here the sum-pooling propagation rule is applied instead of the original GCN~\cite{gcn} to better distinguish local structures. In fact, more powerful GNNs such as Graph Isomorphic Network~\cite{xu2018how} may further improve the performance. The key idea for GAS is that using a few critical structural features as the guidance information to train the model. The features are extracted in a similar way proposed in ReFeX but aggregated only once, normalized and not binned. With a MLP model as the decoder to approximate the features, i.e., $\hat{\mathbf{F}} = \mathrm{MLP}_{dec}(\mathbf{H})$. The loss function is:
\begin{equation}
\label{eq:gas2}
\begin{aligned}
\mathcal{L} = \left \| \mathbf{F} - \hat{\mathbf{F}} \right \|^2_F
\end{aligned}
\end{equation}

\noindent\textbf{RESD~\cite{zhang2021role}}. RESD also adopts ReFeX~\cite{henderson2011s} to extract appropriate features $\mathbf{F}_{ReFeX}$. It uses a Variational Auto-Encoder~\cite{Kingma2014} architecture to learn the low-noise and robust representations:
\begin{equation}
\begin{aligned}
    \mathbf{Z}_i &= \mathrm{MLP}_{enc}(\mathbf{F}_{i}) \\
    \boldsymbol{\mu}_i &= \mathbf{W}_{\boldsymbol{\mu}}\mathbf{Z}_i + \mathbf{b}_{\boldsymbol{\mu}} \\
    \mathrm{log}(\boldsymbol{\sigma}_i) &=  \mathbf{W}_{\boldsymbol{\sigma}} \mathbf{Z}_i + \mathbf{b}_{\boldsymbol{\sigma}} \\
    \mathbf{H}_i &= \boldsymbol{\mu}_i + \boldsymbol{\sigma}_i \odot \boldsymbol{\epsilon}, \boldsymbol{\epsilon} \sim \mathrm{Gaussian}(\mathbf{0},\mathbf{I}) 
    \\
    \hat{\mathbf{F}}_i &= \mathrm{MLP}_{dec}(\mathbf{H}_{i})
\end{aligned}
\label{eq:resd1}
\end{equation}
The VAE model is trained via feature reconstruction. A degree-guided regularizer Eq.(\ref{eq:drne2}) designed in DRNE~\cite{tu2018deep} is introduced in RESD for preserving topological characteristics. The combined objective is as follows:
\begin{equation}
\mathcal{L} = \left \| \mathbf{F} - \hat{\mathbf{F}} \right \|^2_F + \gamma_{deg}\mathcal{L}_{deg}
    \label{eq:resd2}
\end{equation}

\noindent\textbf{GraLSP}~\cite{jin2020gralsp}. GraLSP is a GNN framework integrating local structural patterns that can be employed on role-oriented tasks. For a node $v_i$, it captures structural patterns by generating $w$ random walks starting from $v_i$ with length $l_w$: $\mathcal{W}_i = \{\omega_{i1},...,\omega_{iw}\}$, and then anonymizes them $\mathrm{aw}(\omega)$~\cite{ivanov2018anonymous}. Each anonymous walk $\mathrm{aw}(\omega)$ is represented as an embedding lookup table $\mathbf{u}_{\mathrm{aw}(\omega)}$. Then the aggregation of neighborhood representation is designed as follows:
\begin{equation}
\label{eq:gralsp1}
\begin{aligned}
(\mathbf{H}_{nei})^{(l)}_i &= \mathrm{MEAN}_{\omega \in \mathcal{W}_i, j \in \left [1,\left\lfloor \frac{2l_w}{|\omega|} \right \rfloor\right]}(\alpha_{i,\omega}^{(l)}(\mathbf{a}_{i,\omega}^{(l)}\odot \mathbf{H}_{\omega_j}^{(l-1)}))\\
\mathbf{H}^{(l)}_i&=\mathrm{ReLU}(\mathbf{W}_{self}^{(l)}\mathbf{H}^{(l-1)}_i + \mathbf{W}_{nei}^{(l)}(\mathbf{H}_{nei})^{(l)}_i)
\end{aligned}
\end{equation}
where $\mathbf{W}_{self}$ and $\mathbf{W}_{nei}$ are trainable parameter matrices. $\alpha_{i,\omega}^{(l)}$ is learned attention values based on their local structure:
\begin{equation}
\label{eq:gralsp2}
\begin{aligned}
\alpha_{i,\omega}^{(l)} = \frac{\mathrm{exp}(\mathrm{SLP}_{att}(\mathbf{u}_{\mathrm{aw}(\omega)}))}{\sum_{\omega' \in \mathcal{W}_i}\mathrm{exp}(\mathrm{SLP}_{att}(\mathbf{u}_{\mathrm{aw}(\omega')}))}
\end{aligned}
\end{equation}
$\mathrm{SLP}(\cdot)$ denotes a single-layer perceptron. $\mathbf{a}_{i,\omega}^{(l)}$ is the amplification coefficients:
\begin{equation}
\label{eq:gralsp3}
\begin{aligned}
\mathbf{a}_{i,\omega}^{(l)} = \mathrm{SLP}_{amp}(\mathbf{u}_{\mathrm{aw}(\omega)})
\end{aligned}
\end{equation}
To preserve proximities between nodes, the loss function in DeepWalk~\cite{perozzi2014deepwalk} is leveraged:
\begin{equation}
\label{eq:gralsp4}
\begin{aligned}
\mathcal{L}_{prox} = - \sum_{v_i \in \mathcal{V}} \sum_{v_j \in \mathcal{N}_i} (\mathrm{log}\sigma(\mathbf{H}_i\mathbf{H}_i^\top)\\-\gamma_{neg}\mathbb{E}_{v_k \sim P_n(v)}\left[ \mathrm{log}\sigma(\mathbf{H}_i\mathbf{H}_k^\top) \right])
\end{aligned}
\end{equation}
After $L$ aggregations, the embeddings are $\mathbf{H} = \mathbf{H}^{(L)}$. To capture structural similarities between nodes, GraLSP designs the following loss:
\begin{equation}
\label{eq:gralsp5}
\begin{aligned}
\mathcal{L}_{struc} = &- \sum_{v_i \in \mathcal{V}, \omega_j,\omega_k,\omega_s \in \mathcal{W}_i} \mathrm{log}\sigma(\mathbf{u}_j^\top\mathbf{u}_k-\mathbf{u}_k^\top\mathbf{u}_s),\\
s.t. \;\; &\hat{p}(\omega_j|v_i) > \hat{p}(\omega_j|\mathcal{G}), \hat{p}(\omega_k|v_i) > \hat{p}(\omega_k|\mathcal{G}), \\
  &\hat{p}(\omega_n|v_i) < \hat{p}(\omega_n|\mathcal{G}).&
\end{aligned}
\end{equation}
where $\hat{p}(\cdot)$ is the empirical distribution of anonymous walks:
\begin{equation}
\label{eq:gralsp6}
\begin{aligned}
\hat{p}(\omega_j|v_i) &= \frac{\sum_{\omega\in \mathcal{W}_i} \mathbb{I}(\mathrm{aw}(\omega)=\omega_j)}{w}\\
\hat{p}(\omega_j|\mathcal{G}) &= \frac{\sum_{i=1}^{n}\hat{p}(\omega_j|v_i) }{n}
\end{aligned} 
\end{equation}
The objective is to combine the two losses with a trade-off parameter $\gamma_{struc}$:
\begin{equation}
\label{eq:gralsp7}
\begin{aligned}
\mathcal{L}= \mathcal{L}_{prox} +\gamma_{struc}\mathcal{L}_{struc}
\end{aligned} 
\end{equation}

\begin{table*}[htbp]
\caption{Statistic of the networks for node classification and clustering.}
\centering
\small
\begin{tabular}{|c|c|c|c|c|c|c|c|}
\hline
Dataset & \# Nodes  &  \# Edges   &  \# Classes  &  Density(\%) & Mean Degree  &  Average CC & Transitivity \\
\hline
Brazil &  $131$ &  $1,074$ & $4$ & $12.6130$ & $16.3969$ & $0.6364$ & $0.4497$ \\
Europe & $399$ &  $5,995$ & $4$ & $7.5503$ & $30.0501$ & $0.5393$  & $0.3337$ \\
USA & $1,190$ &  $13,599$ & $4$ & $1.9222$ & $22.8555$ & $0.5011$ & $0.4263$ \\
Reality-call & $6,809$  &  $7,697$  &$3$ & $0.0332$ & $2.2608$ & $0.0178$  & $0.0024$ \\
Actor & $7,779$  &  $26,733$  &$4$ & $0.0888$ & $6.8917$ & $0.0790$  & $0.0156$ \\
Film & $27,312$  &  $122,706$  &$4$ & $0.0329$ & $8.9855$ & $0.1180$  & $0.0278$ \\
\hline
\end{tabular}
\label{tab.datasets1}
\end{table*}

\begin{table*}[htbp]
\caption{Statistic of the networks for top-k similarity search.}
\centering
\small
\begin{tabular}{|c|c|c|c|c|c|c|c|c|}
\hline
Dataset & \# Nodes  &  \# Edges  & \# Bots & \# Admins  &  Density(\%) & Mean Degree  &  Average CC & Transitivity \\
\hline
ht-wiki-talk & $446$  &  $758$  &$24$ & $0$ & $0.7638$ & $3.3991$ & $0.0941$  & $0.0085$ \\
br-wiki-talk &  $1,049$ &  $2,330$ & $35$ & $8$ & $0.4239$ & $4.4423$ & $0.1998$ & $0.0410$ \\
cy-wiki-talk & $2,101$ &  $3,610$ & $31$ & $16$ & $0.1636$ & $3.4365$ & $0.1579$  & $0.0090$ \\
oc-wiki-talk & $3,064$  &  $4,098$  &$43$ & $4$ & $0.0873$ & $2.6749$ & $0.0994$  & $0.0023$ \\
eo-wiki-talk & $7,288$ &  $14,266$ & $120$ & $21$ & $0.0537$ & $3.9149$ & $0.1206$ & $0.0085$ \\
gl-wiki-talk & $7,935$  &  $19,887$  &$12$ & $14$ & $0.6318$ & $5.0125$ & $0.4913$  & $0.0037$ \\
\hline
\end{tabular}
\label{tab.datasets2}
\end{table*}

\noindent\textbf{GCC~\cite{qiu2020gcc}}. GCC is a pre-train Graph Neural Network model, which represents a node as $\mathbf{H}_i$ by encoding its node-centric subgraph and tries to distinguish the similar subgraphs from the dissimilar ones. It aims to leverage several large-scale networks to train the ability of a GNN-based model to discriminate node substructures in an unsupervised manner. When the scale of the pre-trained datasets is large enough, GCC has the power to recognize local connective patterns of nodes, and the representations generated by it can measure structural similarities. While other GNN methods (such as~\cite{You.2021.arXiv},~\cite{you2019position}) either concentrates on community and proximity, or trains model in a supervised manner. So we introduce GCC as role-oriented network embedding method. Specifically, for each node $v_i$ in a network, GCC extracts its $k$-hop reachable neighborhood as $\mathcal{G}_i^k$,  and leverages the Graph Isomorphic Network~\cite{xu2018how} as the encoder whose convolutional layer is:
\begin{equation}
\mathbf{H}^{(l)} = \mathrm{MLP}_{GIN}((\mathbf{A} + (1+\epsilon)\cdot\mathbf{I})\mathbf{H}^{(l-1)})
    \label{eq:gcc1}
\end{equation}
where $\epsilon$ could be a learnable or fixed parameter and the input attributes of GIN is initialized as the eigenvectors of $\mathcal{G}_i^k$. A similar subgraph instance of $v_i$ is induced based on the nodes in the random work with restart starting from $v_i$. $K$ dissimilar instances are subgraphs induced in the same way but starting from other nodes which could be in other networks. The representations $\{ \mathbf{x}_0, ...,\mathbf{x}_K \}$ of these $K+1$ instances are generated via another GIN encoder. The representation of the similar instance is denoted as $\mathbf{x}^+ \in \{ \mathbf{x}_0, ...,\mathbf{x}_K \}$, and GCC is pre-trained by a contrastive learning method called InfoNCE~\cite{oord2018representation}:
\begin{equation}
    \mathcal{L} = \sum_{v_i \in \mathcal{V}} - \mathrm{log}\frac{\mathrm{exp}(\mathbf{H}_i \mathbf{x}^+/\iota)}{\sum_{j=0}^{K} \mathrm{exp}(\mathbf{H}_i \mathbf{x}_j/\iota) }
    \label{eq:gcc2}
\end{equation}
where $\iota$ is a hyper-parameter.

\noindent\textbf{\textit{Remark.}} These deep methods incorporate model traditional role-related concepts used in shallow methods 
with deep learning techniques to map structural information into non-linear latent representations. DRNE learns regular equivalence and reconstructs node degrees. GAS uses structural features to guide the training of graph convolutional networks. RESD combines variants of DRNE and ReFeX via a VAE architecture. GraLSP reconstructs the similarities based on anonymous walks. And GCC is a pre-train model which encodes node-centric subgraphs and is pre-trained via a constrastive learning manner. 
\section{Experimental Evaluation}\label{sec:Experiment}

In this section, we show the comprehensive analysis of these popular role-oriented embedding methods on widely used benchmarks. The experimental evaluations are conducted from the perspectives of both efficiency and effectiveness. To analyze the efficiencies of these methods, we compare their running time for generating node representations on both real-world and synthetic networks with varying sizes. To evaluate the effectiveness of these methods, we select four tasks for the evaluation including (1) the classification experiment based on the ground-truth labels of datasets by comparing the Micro-F1 and Macro-F1 scores, (2) the clustering experiment by comparing some clustering indices with K-Means model in an unsupervised manner. (3) the visualization experiment by plotting the node representations in a $2$-$D$ space to observe the relationships between node embeddings and their roles. (4) the top-k similarity search experiment to see if nodes in the same role are mapped into close position in the embedding space.


It's worth noting that we do not select link prediction for the experimental study, mainly because (1) only limited previous role-oriented NE methods evaluate the performance of this task (see Table~\ref{tab.methodlist}). Thus, it is difficult to make a consistent and fair comparison for all these methods. (2) It has been demonstrated that global information, i.e., roles, is less useful than local information, i.e., proximity, in link prediction task~\cite{lyu2017enhancing}. So we believe that this task is more suitable for proximity-preserving rather than role-oriented node embeddings.

\subsection{Datasets}

We perform our experiments on the following datasets which form unweighted and undirected networks to illustrate the potential of role-oriented network embedding methods in capturing roles. Based on the type of conducted experiments, these networks are divided in two groups. One group of networks, which include \textbf{Brazil}, \textbf{Europe}, \textbf{USA}, \textbf{Reality-call}, \textbf{Actor} and \textbf{Film}, are for node classification and clustering in which the nodes are labeled based on some role-related rules. The other networks are all Wiki-talk networks in which some nodes have the same but very rare role. These Wiki-talk networks are for top-k similarity search. We show some statistical characteristics from various aspects of these networks in Table~\ref{tab.datasets1} and Table~\ref{tab.datasets2}, respectively. More details about the datasets can be found in the Supplementary Materials.

\begin{figure*}[htbp]
    \centering
    {\includegraphics[width=0.99\linewidth]{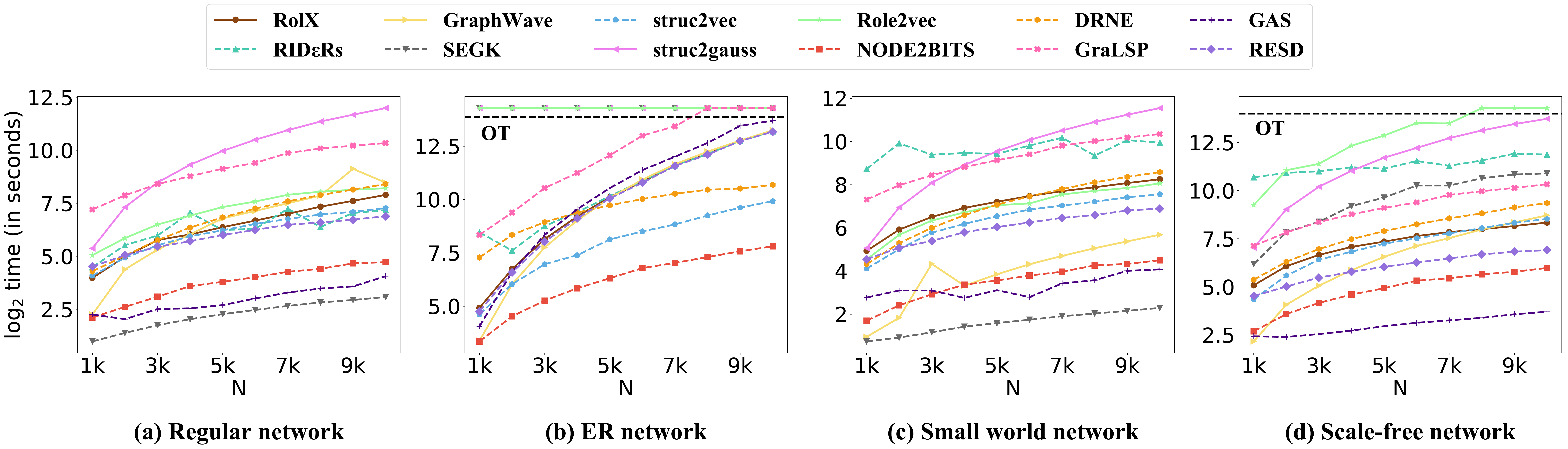}}
    \caption{Running time of $12$ methods on four different types of synthetic networks with different sizes. }
\label{fig:linear time}
\end{figure*}

 \begin{figure}[htbp]
    \centering
    {\includegraphics[width=0.8\linewidth]{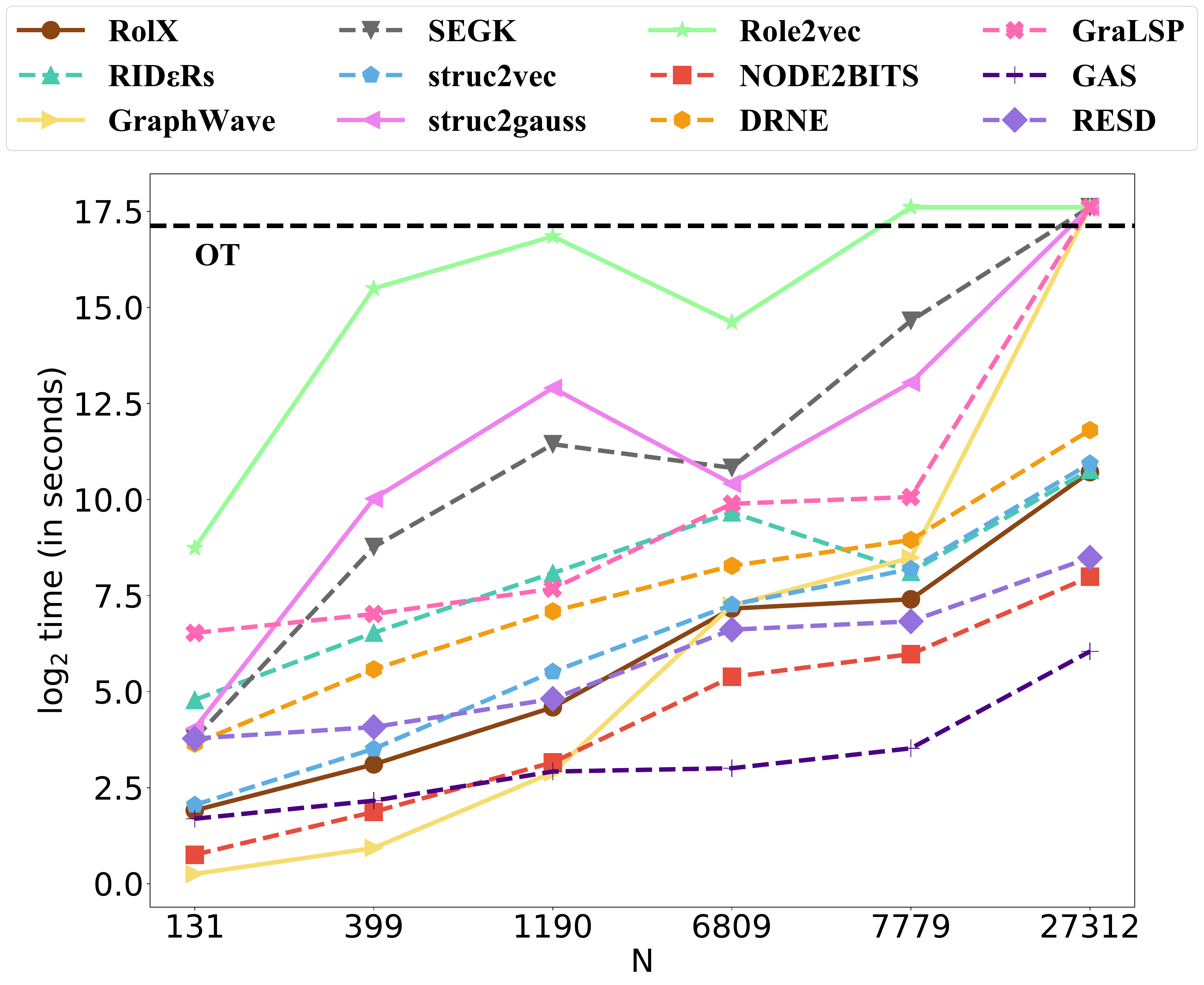}}
    \caption{Running time of $12$ methods on the six real networks. }
\label{fig:time}
\end{figure}

\subsection{Experimental Settings}
Although different methods with different embedding mechanism have been proposed, according to our proposed two-level classification taxonomy discussed in Section~\ref{sec:Definition}, for each class of role-oriented network embedding, we choose $4$ methods to analyze the performance on different role related tasks. In specific, RolX~\cite{henderson2012rolx}, RID$\large\boldsymbol{\varepsilon}$Rs~\cite{gupte2017role}, GraphWave~\cite{donnat2018learning} and SEGK~\cite{nikolentzos2019learning} belong to the low-rank matrix factorization.
struc2vec~\cite{ribeiro2017struc2vec}, struc2gauss~\cite{pei2020struc2gauss}, Role2Vec~\cite{ahmed2019role2vec} and NODE2BITS~\cite{jin2019node2bits} are all based on random walk.
DRNE~\cite{tu2018deep}, GraLSP~\cite{jin2020gralsp}, GAS~\cite{guo2020role} and RESD~\cite{zhang2021role} pertain to the scope of deep learning. In the subsequent experiments, all the parameters are fine-tuned. Note that for Role2vec, we use motif-count features as the default setting, but it causes the out-of-memory issue in large networks and we use node degrees to circumvent. We release the datasets and source code used in the experiments on Github \url{}.

\begin{figure*}[htbp]
    \centering
    {\includegraphics[width=\linewidth]{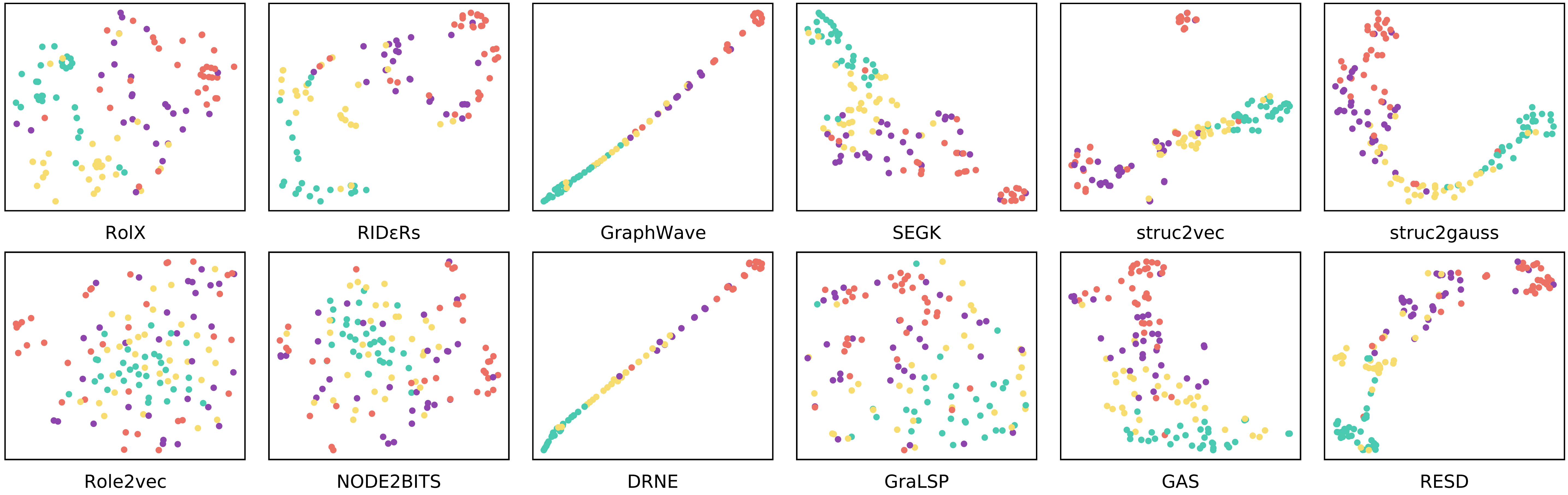}}
    \caption{Visualization results of $12$ methods using t-SNE on the Air Brazil network. The points denote the nodes and the same color indicates that they belong to the same label.Two nodes are more closer to each other if their embeddings are role similar.}
\label{fig:tsne}
\end{figure*}
\begin{table*}[htbp]
    \centering
    \caption{Node classification average F1 score on different networks.}
    \begin{tabular}{|c|c|c|c|c|c|c|c|c|c|c|c|c|}
    \hline
    \multirow{2}{*}{Method} & \multicolumn{2}{c|}{Brazil} & \multicolumn{2}{c|}{Europe} & \multicolumn{2}{c|}{USA} & \multicolumn{2}{c|}{Reality-call} & \multicolumn{2}{c|}{Actor} &
    \multicolumn{2}{c|}{Film} \\
    \cline{2-13}
      & Micro & Macro & Micro & Macro & Micro & Macro & Micro & Macro & Micro & Macro & Micro & Macro \\
    \hline
    RolX & $0.749$ & $0.741$ & $0.556$ & $0.546$ & $0.623$ & $0.617$ & $0.593$ & $0.383$ & $0.465$ & $0.451$ & $\textbf{0.494}$ & $\textbf{0.396}$ \\
    RID$\large\boldsymbol{\varepsilon}$Rs & $\textbf{0.790}$ & $\textbf{0.783}$ & $0.539$ & $0.519$ & $0.626$ & $0.618$ & $0.635$ & $0.396$ & $\textbf{0.470}$ & $0.448$ & $0.479$ & $0.380$ \\
    GraphWave & $\textbf{0.762}$ & $\textbf{0.753}$ & $0.526$ & $0.491$ & $0.517$ & $0.469$ & $\textbf{0.839}$ & $\textbf{0.516}$ & $0.251$ & $0.180$ & OM & OM \\
    SEGK & $0.723$ & $0.718$ & $0.536$ & $0.524$ & $0.615$ & $0.607$ & $\textbf{0.839}$ & $\textbf{0.514}$ & $\textbf{0.479}$ & $\textbf{0.460}$ & OM & OM \\
    \hline
    struc2vec & $\textbf{0.766}$ & $\textbf{0.757}$ & $\textbf{0.577}$ & $\textbf{0.572}$ & $0.599$ & $0.594$ & $0.593 $ & $0.376$ & $0.463 $ & $\textbf{0.456}$ & $0.474$ & $0.365$ \\
    struc2gauss & $0.730$ & $0.715$ & $\textbf{0.585}^{\star}$ & $\textbf{0.580}^{\star}$ & $\textbf{0.641}$ & $\textbf{0.632}$ & $0.603$ & $0.391$ & $0.456$ & $0.449$ & OT & OT \\
    Role2vec & $0.385$ & $0.358$ & $0.362$ & $0.347$ & $0.467$ & $0.455$ & $0.541$ & $0.354$ & $0.277\dag{}$ & $0.277\dag{}$ & $0.278\dag{}$ & $0.243\dag{}$ \\
    NODE2BITS & $0.593$ & $0.577$ & $0.490$ & $0.477$ & $0.583$ & $0.573$ & $0.524$ & $0.354$ & $0.442$ & $0.424$ & $\textbf{0.542}^{\star}$ & $\textbf{0.428}^{\star}$ \\
    \hline
    DRNE & $0.716$ & $0.700$ & $0.542$ & $0.521$ & $0.601$ & $0.588$ & $0.630$ & $0.478$ & $0.462$ & $0.449$ & $0.453$ & $0.339$ \\
    
    GAS & $0.757$ & $0.750$ & $\textbf{0.583}$ & $\textbf{0.574}$ & $\textbf{0.668}^{\star}$ & $\textbf{0.659}^{\star}$ & $\textbf{0.841}^{\star}$ & $\textbf{0.529}^{\star}$ & $\textbf{0.480}^{\star}$ & $\textbf{0.466}^{\star}$ & $\textbf{0.490}$ & $\textbf{0.380}$ \\
    RESD & $\textbf{0.797}^{\star}$ & $\textbf{0.791}^{\star}$ & $0.557$ & $0.544$ & $\textbf{0.640}$ & $\textbf{0.634}$ & $0.607$ & $0.411$ & $\textbf{0.476}$ & $\textbf{0.463}$ & $\textbf{0.516}$ & $\textbf{0.415}$ \\
    GraLSP & $0.510$ & $0.490$ & $0.455$ & $0.422$ & $0.535$ & $0.523$ & $0.454$ & $0.300$ & $0.341$ & $0.316$ & OM & OM \\
\hline
\end{tabular}
\label{tab.3}
\end{table*}

\subsection{Efficiency analysis}

All experiments are performed on a machine with Intel(R) Xeon(R) CPU E5-2680 v4 at 2.40GHz and 125GB RAM. 
In this experiment, by ignoring the differences in some implementation details, we report the running time of these methods on the above six real networks. The results are shown in Fig~\ref{fig:time}. The y-axis represents the average logarithmic time (seconds) of $10$ times running of each method on one specific network and x-axis is the size (number of nodes) of the network. If the value is above the dotted line, it means that the cost of corresponding method is beyond our tolerance.
Generally speaking, the methods based on deep learning are relatively efficient compared to others, especially for the methods that need higher-order features. GAS~\cite{guo2020role}, RESD~\cite{zhang2021role} and NODE2BITS~\cite{jin2019node2bits} are the three most efficient methods and RolX~\cite{henderson2012rolx} also has competitive results.  struc2gauss~\cite{pei2020struc2gauss}, SEGK~\cite{nikolentzos2019learning} and Role2vec~\cite{ahmed2019role2vec} cost more time to learn the node embeddings since they need to compute higher-order features, e.g., motif. In particular, as the most classical method proposed in the early stage, RolX still shows competitive in efficiency.

However, the numbers of nodes in these networks are quite different in a random way. In order to show the efficiency of different methods more fairly, we generate four different types of networks with linear variation of the size. The details are as following. 
Regular network. It is a regular graph in which each node has the same number of neighbors and we set it as $3$.
ER network~\cite{erdos59a}. It is based the ER random graph and we set the probability of having a link between any two nodes as $0.1$.
Small world network. It comes from the popular Watts–Strogatz model~\cite{watts1998collective} and for each link, we set it rewiring probability as $0.5$.
Scale-free network. It is based on the Barabási–Albert model~\cite{barabasi1999emergence} with  preferential attachment, for each new nodes, the number of its links is set as $3$.

For each type of network, we vary the number of nodes $N$ from $1,000$ to $10,000$. The running time of the baselines is shown in Fig~\ref{fig:linear time}. From these results, some conclusions can be drawn. Firstly, the ER network is relatively dense for its fixed link probability. The average degree of nodes in larger networks is larger correspondingly. In this type of network, NODE2BITS has significant efficiency advantage than all others, and it also achieves competitive results in other three types of networks. Secondly, SEGK is very efficient on the regular and small world networks but time-consuming on the ER and scale-free networks. Thirdly, GAS is the most effective on the scale-free networks and has second results only to the SEGK. Besides, RolX and RESD rank in the middle of this experiments for they all are based on ReFex~\cite{henderson2011s} and reconstruct the features with different methods. GAS is one of the most efficient on almost all the networks. GAS reconstructs a small number of primary features which leads to few parameters and epochs for loss convergence. NODE2BIT consumes little time on all the networks stably. Because its feature aggregation process, which is based on random walk and the Simhash mapping process, is almost linear to the number of edges.


\begin{figure*}[htbp]
    \centering
    {\includegraphics[width=0.8\linewidth]{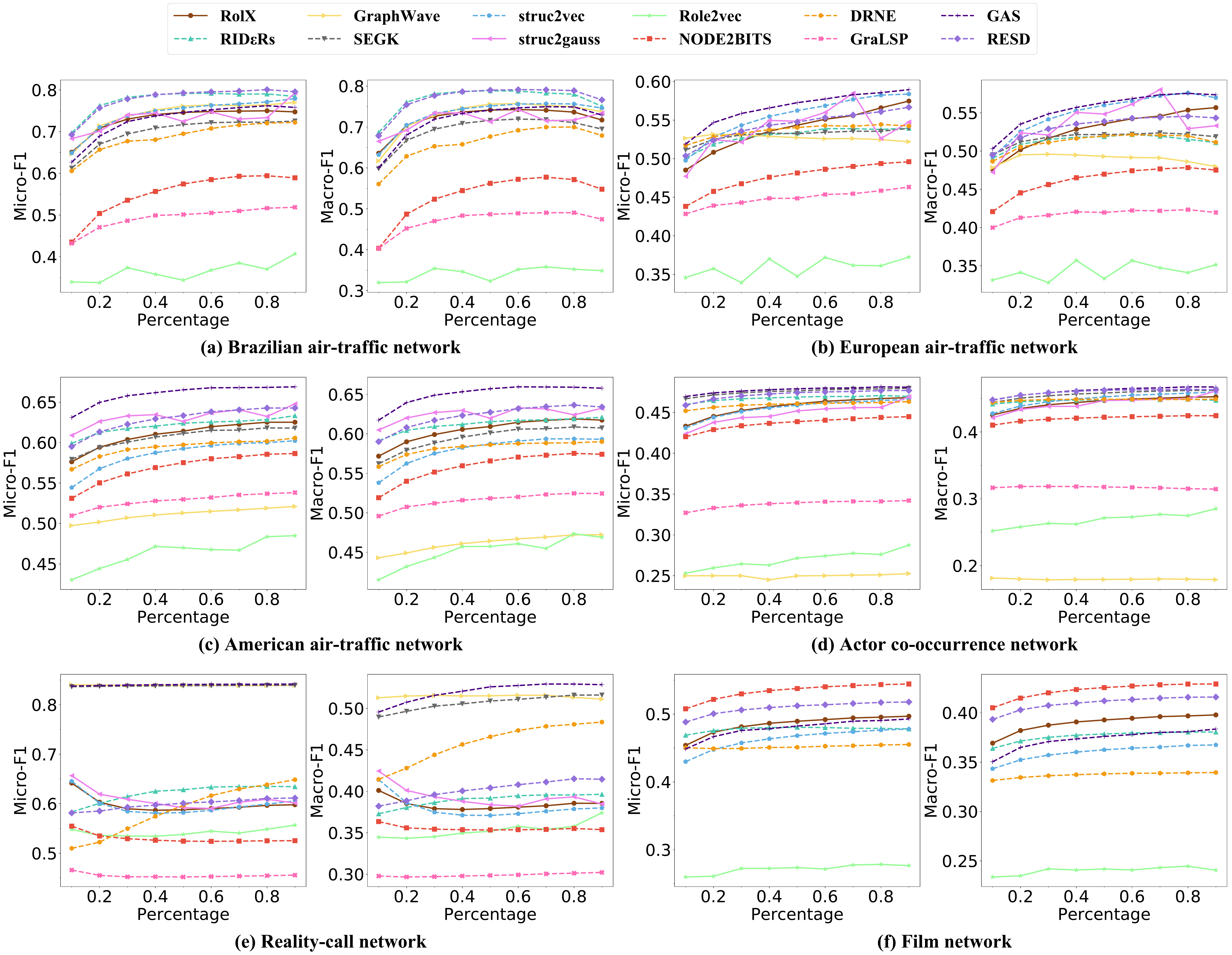}}
    \caption{Classification results of the $12$ role-oriented network embedding methods on the six real world networks.The performance is based on the Micor-F1 and Macor-F1 with different percentages of training set, which is the average values of $20$ runs.}
\label{fig:classification}
\end{figure*}

\subsection{Visualization analysis}
Visualization can help to understand the performance of the different methods intuitively. It is also a common way for understanding the network structure and the methods. In this section, we show the t-SNE results of these methods based on the Brazil network in Fig~\ref{fig:tsne}. It can be observed that, RID$\large\boldsymbol{\varepsilon}$Rs~\cite{gupte2017role}, GraphWave~\cite{donnat2018learning}, SEGK~\cite{nikolentzos2019learning}, struc2vec~\cite{ribeiro2017struc2vec}, struc2gauss~\cite{pei2020struc2gauss}, DRNE~\cite{tu2018deep}, GAS~\cite{guo2020role} and RESD~\cite{zhang2021role} can be more accurate to identify the labels compared with others. The cyan nodes correspond to the airports with large passenger flow and all these methods can embed it effectively except NODE2BITS~\cite{jin2019node2bits}. The red nodes represent the marginal airports. These nodes are usually hard to be well identified for they are more scattered in the network. Therefore, these methods can not identify this kind of nodes well. Besides, although we expect to embed nodes with same role into close points in latent space, it is also still the demand nodes should also be slightly different even they are in the same role. From this view, struc2gauss, SEGK and GAS are better selections. GAS benefits from the low-dimensional features which leads to the same-role embeddings gathering. The embeddings of SEGK and struc2gauss shape some clumps since they are directly constrained by effective similarities.

\subsection{Classification results}
Here we report the performance of the classification experiment on the six networks. 
In detail, with the learned node embeddings of each method, we randomly select $70\%$ of the nodes as the training set and the others as the test set and take the logistic regression classifier to learn the labels of nodes. The results are shown in Table~\ref{tab.3}. It represents the Micro-F1 and Macro-F1 value of classification results. To avoid losing of generality, we report the average results of $20$ times of independent runs on each network.

As shown in Table~\ref{tab.3}, for each column, we mark the values of methods with significant advantages, i.e. the top results of these methods. OM and OT mean that it cannot be calculated for fixed memory and limited time, and the \dag{} denotes that Role2vec uses degree features rather than motif features because it is very time-consuming. 

\begin{figure*}[htbp]
    \centering
    {\includegraphics[width=0.8\linewidth]{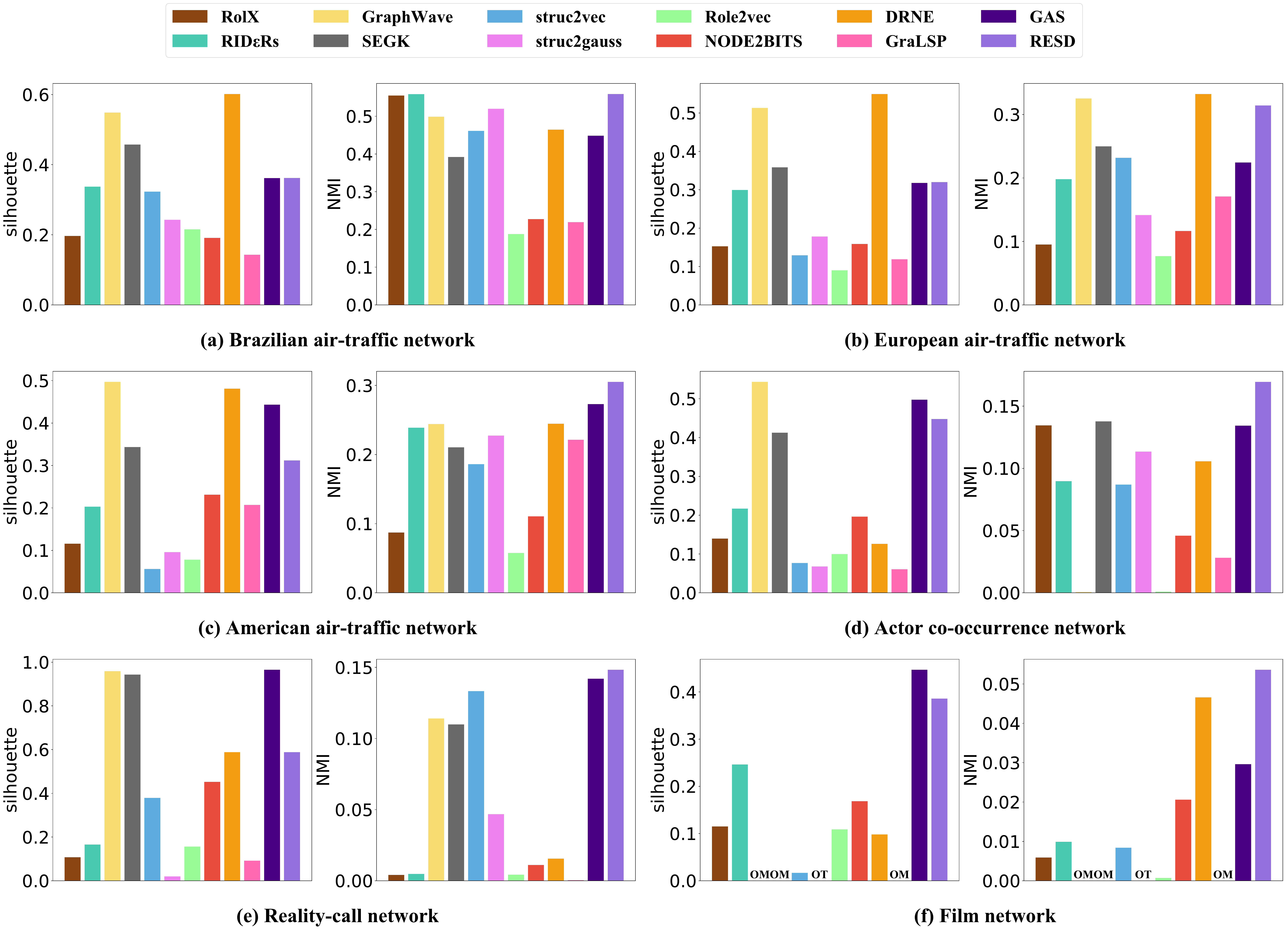}}
    \caption{Clustering results of the $12$ role-oriented network embedding methods on the six real world networks. The performance is based on the silhouette coefficient and NMI and evaluated on $10$ times run.}
\label{fig:clustering}
\end{figure*}

For the three air-traffic networks, RESD~\cite{zhang2021role} performs best on Brazil and  RID$\large\boldsymbol{\varepsilon}$Rs~\cite{gupte2017role} follows because the ReFeX~\cite{henderson2011s} features used by them can effectively capture structural similarities in small networks. While on the other two networks, GAS~\cite{guo2020role} and struc2gauss~\cite{pei2020struc2gauss} perform well. For the rest three large networks, struc2guass cannot generate node embeddings within limited time, though it performs well in small networks. GAS outperforms others on Reality-call and Actor, while NODE2BITS~\cite{jin2019node2bits} gets the highest score on Film.

Among all the datasets, we observe that Role2vec~\cite{ahmed2019role2vec} obtains the worst performance. This may because that it leverages the features-based random walk and damages the original network structure. It can be applied to link prediction task, but is not proper for role discovery. Though GraLSP~\cite{jin2020gralsp} leverages the local structural patterns, it concentrates more on node proximity and fails to distinguish roles of nodes.

In Section~\ref{sec:algorithm}, we have divided these methods into three types: random walk, matrix factorization, and deep learning. In general, deep learning methods perform well on all the networks. For the three small air-traffic networks, random walks (struc2vec~\cite{ribeiro2017struc2vec} and struc2gauss) perform well, while matrix factorization methods get better results on large networks, even if some of them cannot obtain embeddings because of the out of memory error. Overall, deep learning methods are scale-well and can effectively discover roles.

For matrix factorization methods, RID$\large\boldsymbol{\varepsilon}$Rs is better than RolX~\cite{henderson2012rolx} except on Film, and they both can be applied to large-scale networks. GraphWave~\cite{donnat2018learning} and SEGK~\cite{nikolentzos2019learning} perform best on Reality-call but consume too much memory. For random walk methods, struc2vec and struc2gauss show their superiority on small networks, while NODE2BITS is more suitable for large-scale networks. As for deep learning methods, GAS and RESD stay ahead of other approaches on all networks.

As shown in Fig.~\ref{fig:time}, for RESD, NODE2BITS, and GAS, there is a linear relationship between their computational complexities and network sizes. In Table~\ref{tab.3}, GAS and RESD perform well on all networks, while NODE2BITS only gets the best score on Film. It is obvious that combining structural features and deep learning models has advantages in both time and classification.

Further more, in order to analyze the influence of different percentages of training set on classification results, we also show the performance of these methods with varying size of training (from $10\%$ to $90\%$). The results are shown in Fig~\ref{fig:classification}. The experimental settings are same introduced above. 

In general, as the ratio of the training set increases, the F1 scores of classification results are improved. On the three air-traffic networks, we observe significant increase, but when we use too many training samples (more than $80\%$), the scores go down because of over-fitting. The scores on large networks are stable, because small percentage of training samples are sufficient enough to make the linear model constraint.

The scores on Reality-call network are different to others, because the distribution of labels is unbalance, and the Micro-F1 scores are higher than Macro-F1. In general, we observe the scores of Role2vec fluctuate, which means that Role2vec fails for the task of role classification. GraLSP also performs a litter better than Role2vec, but it gets worse on Reality-call. Other methods show similar tendencies, while deep learning methods (GAS and RESD) and NODE2BITS exhibit significant superiority.

\subsection{Clustering results}
In the social science, the role discovery is usually denoted as an unsupervised task which can be implemented by the unsupervised clustering in machine learning. In this experiment, we study the role clustering problem based on the embedding of different methods. Based on the embedding, we take the k-means algorithm to obtain the clusters, and report the silhouette coefficient~\cite{rousseeuw1987silhouettes} and Normalized Mutual Information (NMI)~\cite{danon2005comparing} in Fig.~\ref{fig:clustering}. The silhouette coefficient is a measure of how similar an object is to its own cluster compared to other clusters. The result only relies on the embeddings themselves rather than true labels. While the NMI measures the similarity between clustering results and real distribution.

We observe that GraphWave~\cite{donnat2018learning} obtains high scores on both measurements in air-traffic networks, but in larger networks (Actor and Reality-call), it gets low NMI scores. DRNE~\cite{tu2018deep} shows similar tendency because its aggregation methodology leads to obvious clustering, but may not fit the true distributions. The results of Role2vec~\cite{ahmed2019role2vec} and GraLSP~\cite{jin2020gralsp} are similar to classification, and they fail to cluster roles of nodes. The performances of other random walk and matrix factorization methods are moderate. In general, it can be observed that deep learning methods (GAS~\cite{guo2020role} and RESD~\cite{zhang2021role}) show overall superiority. They perform well on small networks but significantly outperform others in large-scale networks. This is also consistent with findings of deep learning methods in other domains that deep learning methods can better learn the patterns from larger-scale data.

\subsection{Top-k similarity search}
We conduct top-k similarity search experiments on the $6$ Wiki-talk networks to evaluate the ability to retrieve rare roles of different methods. In specific, after generating role-based embeddings for each node through a candidate method, we find the $k$ most similar users for each bot or administrator by computing the euclidean distance between the embeddings. The mean precision for both bots and administrators on different sizes of the retrieved node list, i.e., different $k$s, is reported in the Supplementary Materials.

In general, we assume that nodes literally having the same role (such as bots in Wiki-networks) in a network have similar structural patterns. However, there are more bots than administrators in most Wiki-talk networks, almost every method achieves higher precision on searching administrators than that on searching bots when the retrieved list size $k$ is fixed and smaller than or close to the number of administrators. This shows that the structures of bots are more irregular or these methods can not effectively capture the specific structural patterns of bots. The structural representation learning on literal roles is much more difficult than on function-based defined roles (such as the classes in Table~\ref{tab.datasets1}), since the latent patterns of literal roles are usually not reflected in the plain structural properties.  

What's more, the inconsistency between literal roles and node structure is obvious across networks. We can come to this conclusion gradually through the following observations: 
the proportion of the same literal role in different networks varies greatly as shown in Table~\ref{tab.datasets2}; none of the compared methods can produce top results on all the networks with $k$ fixed; although the proportion of bots in cy-wiki-talk and oc-wiki-talk networks are very close, the results of almost every method on cy-wiki-talk are better than those on oc-wiki-talk. Thus, the literal roles in different networks are less comparable and transferable to those roles (such as the classes in Table~\ref{tab.datasets1}) defined directly by node functions.

Though no one method achieves best performance on all the networks, the ranks of each method on a network with different retrieved list size $k$ are usually close. Some methods, such as RID$\large\boldsymbol{\varepsilon}$Rs~\cite{gupte2017role}, GraphWave~\cite{donnat2018learning} and NODE2BITS~\cite{jin2019node2bits} show their wider applicability across different networks. NODE2BITS surprisingly achieves good and stable results compared with its performance on node classification and clustering tasks. Its feature aggregation mechanism based on random walks makes it lose some elaborate information but capture more latent structural patterns. Similarly, RID$\large\boldsymbol{\varepsilon}$Rs and GraphWave can also capture more latent structural patterns via $\varepsilon$-equitable refinement and characterizing wavelet distributions, respectively. Role2vec still lacks competitiveness on this task, since it assigns roles based on features firstly and generates embeddings capturing proximities between roles as we argued above. The deep learning methods DRNE~\cite{tu2018deep}, GAS~\cite{guo2020role} and RESD~\cite{zhang2021role} show mediocre performance as they overfit some features due to the deep process of feature reconstruction. While the similarity captured by GraLSP~\cite{jin2020gralsp} based on anonymous walks is too coarse.



\subsection{Discussion on experiments}

In this section, we show a variety of experiments on some popular role-oriented network embedding methods. These methods cover all the three major categories and the five minor categories we propose. We evaluate their effectiveness on classification, clustering, visualization and top-k similarity search and efficiency on networks in different scales.

We find that no one method can outperform the others on all tasks. As we argued above, each task has a different focus and the evaluated methods are only suitable for parts of them. On balance, GAS~\cite{guo2020role} and RESD~\cite{zhang2021role} are good at visualization, node classification and clustering tasks. NODE2BITS~\cite{jin2019node2bits} show outstanding performance on both top-k similarity search and efficiency test. Role2vec~\cite{ahmed2019role2vec} ceases to be effective on these role-based tasks as it assigns roles based on features directly while the embeddings do not capture the structural similarities. It specializes in link prediction for non-direct role-based tasks. Similarity,  the structural similarity objective of GraLSP~\cite{jin2020gralsp} is not powerful enough for role-based tasks. The performance of the other methods fluctuates a lot on different tasks and networks. Thus, it is still an open problem that how we can capture both fine and essential role information effectively and efficiently. 

Though some deep learning methods seems good at many experiments, they have not yet addressed the nature of the role. The deep learning manners give them better capabilities on machine learning tasks such as classification and clustering. However, they utilize additional skills to extract structural information, and lack a suitable objective function to embed it without bias and impurities. Thus, deep role-base representation research, which is still in the beginning stage, is promising and needs more attention. 

\section{Applications}\label{sec:applications}

Role discovery can complement community detection in network clustering and has always been the focus of social science and network science research. Role oriented network embedding has gradually become one of the focuses of graph machine learning. It can benefit to node ranking, community detection, information spreading and other problems in complex networks. It also contributes to the research and development of graph neural networks in machine learning. So we summarize some important applications of role oriented network embedding as follows.

First, it is beneficial to a variety of network mining tasks. Community detection and role discovery are two complementary tasks from local and global perspective of networks structures respectively~\cite{pei2020onlocal}. Thus, combining these two tasks can help to achieve better network generation and mutually improve each other. For instance, MMCR has been proposed to integrate community detection and role discovery in a unified model and detect both of them simultaneously for information networks~\cite{chen2016integrating}. The proposed method extends the Mixed Membership Stochastic Blockmodel (MMSB)~\cite{airoldi2008mixed} to combine the generative process of both community and role. REACT~\cite{pei2019joint} analyzes the community structure and role discovery under a unified framework and describes their relations via non-negative matrix factorization. Moreover, role-oriented embedding can also help other tasks including link prediction~\cite{lyu2017enhancing}, anomaly detection~\cite{henderson2012rolx}, and structural similarity search on network~\cite{rossi2020proximity}.

Second, it can help to analyze the network dynamics and evolution. As we know, nodes with different roles may have inequable influence in information diffusion. RAIN~\cite{wof} analyzes the effect of users with different roles on their reposting messages and models the generation of diffusion process under a unified probabilistic framework. Then, with a general representation learning on the network, it can effectively improve the performance of cascade~\cite{10.1145/3336191.3371811} and its popularity~\cite{10.1109/infocom41043.2020.9155349} prediction. Besides, analyzing the roles of nodes could be used to predict the network evolution and its dynamic behaviors~\cite{7303952,LI2020458} and detecting the varying of roles of node could capture the dynamic structural information~\cite{10.24963/ijcai.2018/531}.

Moreover, it has induced some new graph neural networks with more expressive ability. Although GNNs with the message passing are mainly used for supervised node or graph classification and link prediction, some new architectures have been developed to improve their expressive ability with the heterogeneity and roles in the network. Geom-GCN~\cite{pei2020geom-gcn:} has been proposed to use the role embedding method 
to obtain the structural neighborhood and a bi-level aggregation in GNN. ID-GNN~\cite{You.2021.arXiv} extracts the ego network centered at one node of the network and takes rounds of heterogeneous message passing. It can compute the shortest path and clustering coefficient of the network. These features are are extremely relevant to the role. GCC~\cite{qiu2020gcc} also makes use of ego network and then employs pre-training for role-oriented network embedding.

Besides, it can shed light on new patterns discovery in specific networks. Role discovery and identification can be used to predict the social behaviors, identities and the temporal patterns~\cite{zhang.2017.qi}.~\cite{Zygmunt.2020.Demazeau} analyzes the user behaviors in the Blog social network and learn roles from the concepts of activity, influence and competition. It also can predict the sentiment and topics with the proposed definitions of roles. EMBER~\cite{jin2019smart} considers professional role inference in a large-scale email network, so it can capture and distinguish the behavior similarity of nodes automatically. SADE~\cite{pei2020subgraph} combines role information and subgraph embedding to detect subgraph-level anomalies from financial transaction networks.

\section{Future Directions}\label{sec:direction}
Although many creative and innovative methods have been proposed for role oriented network embedding, it still faces some problems and challenges to be solved in this field. In this section, we will discuss these challenges as future directions.


\textbf{Role-oriented embedding on dynamic networks.}
Most methods reviewed in this survey are for static networks. However, real-world networks are naturally dynamic and continuously streaming over time with evolving structures. Thus role-oriented NE methods for dynamic networks are of fundamentally practical and theoretical importance. There have only been a few approaches for role discovery on dynamic networks e.g., Role-Dynamics~\cite{10.1145/2187980.2188234} and DyNMF~\cite{10.24963/ijcai.2018/531}. However, some questions remain open. For example, how to separate the dynamic networks into different snapshots and how to model the deletion of nodes and edges. Thus, additional investigation is needed to extend current NE methods for dynamic scenarios.


\textbf{Other types of embedding spaces.} Beside to the Euclidean space, a variety of methods project the nodes into the Hyperbolic space~\cite{Peng.2021.Zhao} in proximity preserved network representation learning. For role oriented embedding, as we know, the only work is Hyperboloid~\cite{10.1145/3340531.3412102} which extends the struc2vec~\cite{ribeiro2017struc2vec} with the hyperboloid model. Considering the non-Euclidean space is quite suitable for network embedding because of the power-law distribution of network data, it is interesting and meaningful to have some in-depth analysis and understanding in utilizing hyperbolic neural network for this problem.

\textbf{Construction of larger-scale benchmarks.} All the methods for role oriented network embedding are evaluated on relatively small-scale networks data with thousands of nodes (seen in section~\ref{sec:Experiment}). However, real-world networks are often of a massive scale, e.g., there are billions of users in social networks. Constructing larger-scale benchmark datasets is very important to evaluate existing approaches in effectiveness, efficiency and robustness, and also beneficial for researchers to develop new models.

\textbf{Interpretation on roles and role-oriented embeddings.} In social science, roles often correspond to social identifications, e.g., students and teachers in a school. Real-world networks may not contain such information and thus are difficult to understand the discovered roles. The lack of interpretability of roles and role-oriented embeddings may significantly impact our ability to gain insights into these roles. However, so far, there have not been much work focusing on discovering roles that are interpretable. One exception is RolX~\cite{henderson2012rolx}, and it attempts to explain roles based on several structural measures such as centrality. More research is needed to discover roles and learn embeddings that are more interpretable.

\textbf{Deep theoretical analysis.} Important but not the final, network embedding for role discovery lacks theoretical analysis. Although some methods stem from certain type of equivalence relation, the theoretical analysis has been ignored. Moreover, current methods cannot capture universal node representations and provide the upper/lower bound of the expressive ability. It is expected to have solid theoretical analysis in this filed similar to that in graph neural networks~\cite{xu2018how}.

\section{Conclusions}\label{sec:conclusions}
Role-oriented network embedding can complement the category of network representation learning. It has gradually become one of most important research focuses in network embedding. In this survey, we have proposed a general understanding mechanism for role-oriented network embedding approaches and a two-level classification ontology based on different embedding principles. Using this ontology, we categorize a series of popular methods into different groups. Then, we review the principles and innovations of selected representative embedding methods in different categories. We further conduct comprehensive experiments to evaluate these representative methods, including role classification and clustering, top-k similarity search, visualization and their efficiency test. Last, we summarize the important applications of this problem and outline some future directions. We believe that this survey can help to understand and deepen role-oriented network embedding and it will attract more attention from network science and deep learning.

\section*{Acknowledgment}

This work is supported by the National Natural Science Foundation of China (61902278). 
\ifCLASSOPTIONcaptionsoff
  \newpage
\fi



\bibliographystyle{IEEEtran}
\bibliography{reference}
%

%

\begin{IEEEbiography}[{\includegraphics[width=1.2in,height=1.2in,clip,keepaspectratio]{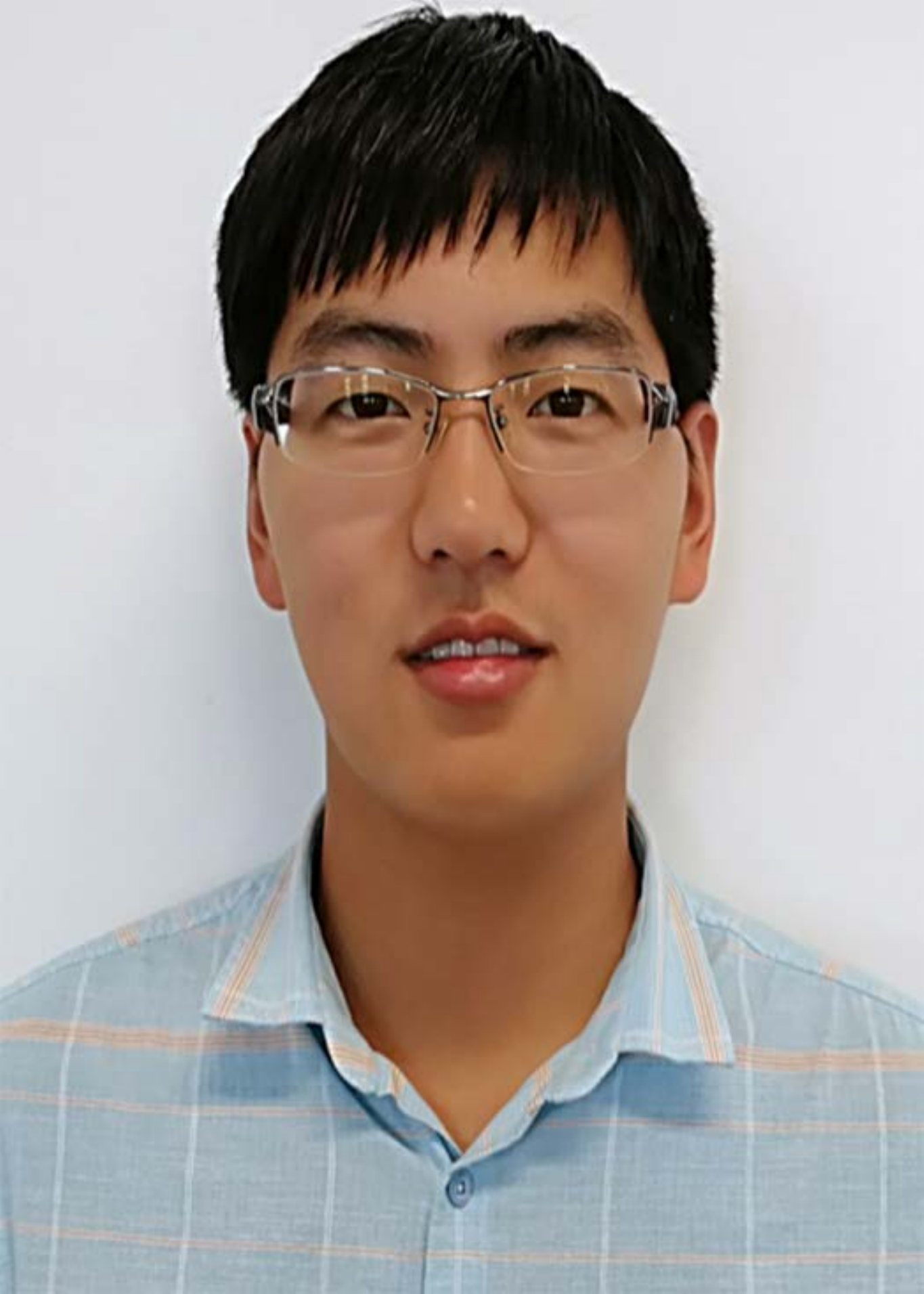}}]{Pengfei Jiao}
 received the Ph.D. degree in computer science from Tianjin University, Tianjin, China, in 2018. He is a lecture with the Center of Biosafety Research and Strategy of Tianjin University. His current research interests include complex network analysis, data mining and graph neural network and applications of statistical network model. He has published more than 50 international journals and conference papers. 
\end{IEEEbiography}
\vspace{-1cm}

\begin{IEEEbiography}[{\includegraphics[width=1in,height=1.25in,clip,keepaspectratio]{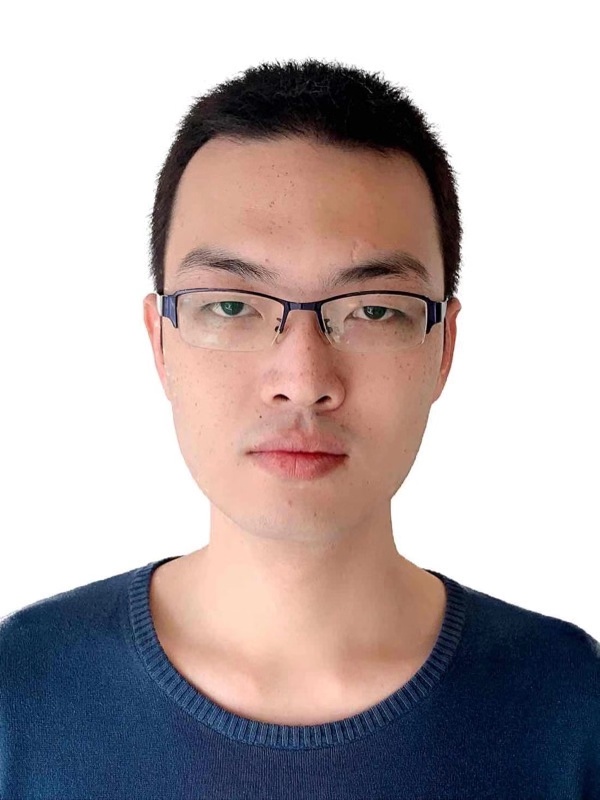}}]{Xuan Guo} is pursuing a doctoral degree at the College of Intelligence and Computing, Tianjin University, P. R. China. His current research interests include complex network analysis, role discovery,  network representation learning and percolation model.
\end{IEEEbiography}
\vspace{-1cm}
 
\begin{IEEEbiography}[{\includegraphics[width=1in,height=1.25in,clip,keepaspectratio]{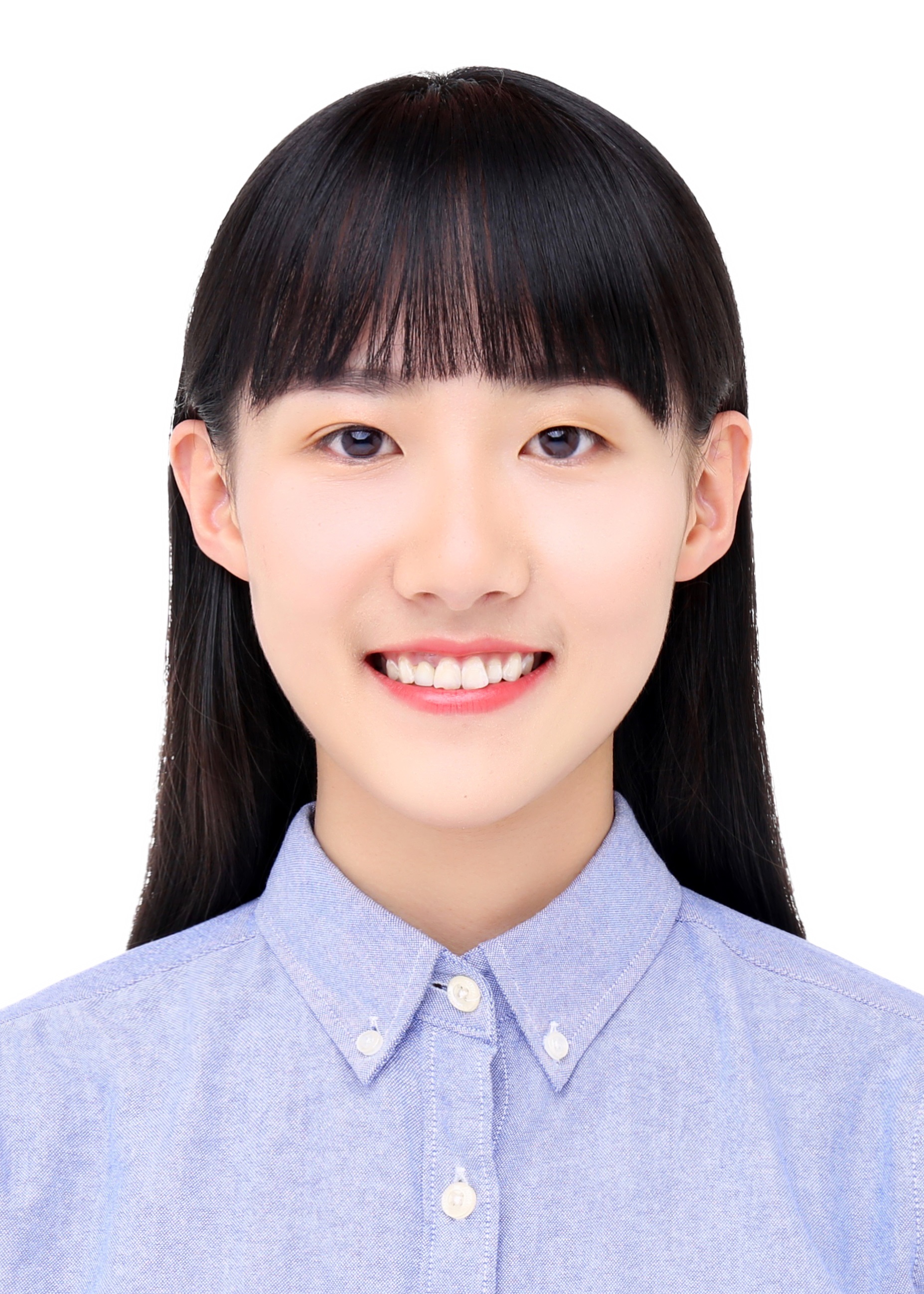}}]{Ting Pan} received the Bachelor degree from Xiamen University in 2020. She is currently pursuing a master's degree at the School of Computer Science and Technology, Tianjin University. Her current research interests include complex network analysis and role-based network representation learning.
\end{IEEEbiography}
\vspace{-1cm}
 
\begin{IEEEbiography}[{\includegraphics[width=1in,height=1.25in,clip,keepaspectratio]{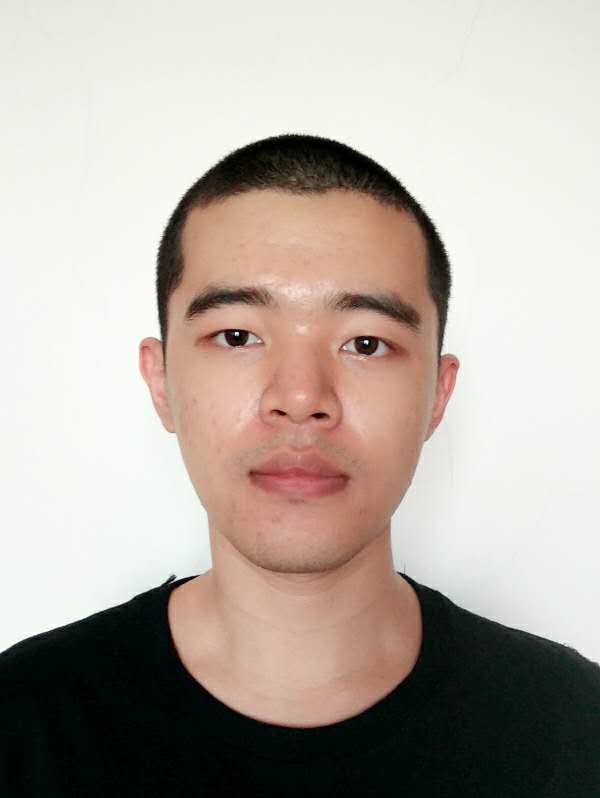}}]{Wang Zhang} received the Bachelor degree from Tianjin University in 2018. He is currently pursuing a master's degree at the School of Computer Science and Technology, Tianjin University. His current research interests include complex network analysis and network embedding.
\end{IEEEbiography}
\vspace{-1cm}

\begin{IEEEbiography}[{\includegraphics[width=1in,height=1.25in,clip,keepaspectratio]{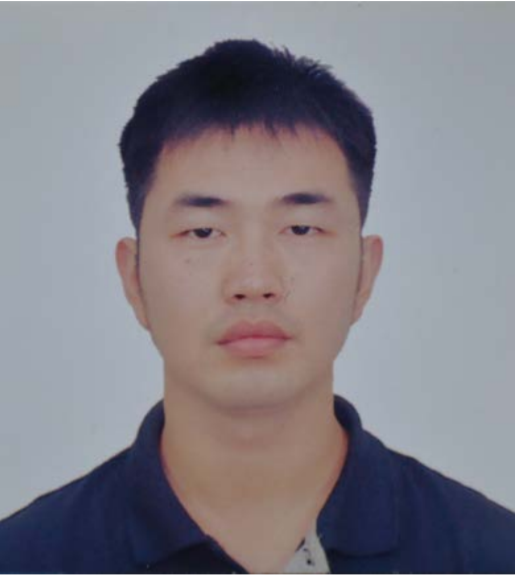}}]{Yulong Pei} received the Ph.D. degree in computer science from Eindhoven University of Technology, Eindhoven, the Netherlands, in 2020. He is a postdoc researcher in the Department of Mathematics and Computer Science, Eindhoven University of Technology. His current research interests include graph mining, social network analysis, and anomaly detection.
\end{IEEEbiography}




\end{document}